\documentclass[10pt,journal,twocolumn]{IEEEtran}
\IEEEoverridecommandlockouts
\usepackage{cite}
\usepackage{amsmath,amssymb,amsfonts,booktabs}
\usepackage{algorithmic}
\usepackage{graphicx}
\usepackage{textcomp}
\usepackage{xcolor}
\usepackage{enumerate}
\usepackage{epstopdf}
\usepackage{subcaption}
\usepackage{amsmath,amsfonts}
\usepackage{amsthm}
\usepackage{amssymb}
\usepackage{easyReview}
\usepackage{algorithmic}
\usepackage{algorithm}
\usepackage{array}
\usepackage[caption=false,font=normalsize,labelfont=sf,textfont=sf]{subfig}
\usepackage{easyReview}
\usepackage{textcomp}
\usepackage{stfloats}
\usepackage{url}
\usepackage{bm}
\usepackage{makecell}
\usepackage{verbatim}
\usepackage{graphicx}
\usepackage{caption}
\usepackage{subfig}
\usepackage{subcaption}
\usepackage{cite}
\usepackage{booktabs}
\usepackage{amsthm}

\newtheorem{proposition}{Proposition}

\usepackage{adjustbox}

\usepackage{array}
\definecolor{mygray}{RGB}{240,240,240}

\usepackage{algorithm}
\usepackage{algorithmic}
\allowdisplaybreaks[4]

\def\BibTeX{{\rm B\kern-.05em{\sc i\kern-.025em b}\kern-.08em
		T\kern-.1667em\lower.7ex\hbox{E}\kern-.125emX}}
\begin{document}

\title{UAV-Enabled ISAC with Fluid Antennas for Low-Altitude Wireless Networks}

\author{Wenchao Liu,
        Xuhui Zhang,
        Jinke Ren,
        Weijie Yuan,     
        Changsheng You,
        and Shuangyang Li

\thanks{
Wenchao Liu and Weijie Yuan are with the School of Automation and Intelligent Manufacturing, Southern University of Science and Technology, Shenzhen 518055, China. (e-mail: wc.liu@foxmail.com; yuanwj@sustech.edu.cn).
}

\thanks{
Xuhui Zhang and Jinke Ren are with the Shenzhen Future Network of Intelligence Institute, the School of Science and Engineering, and the Guangdong Provincial Key Laboratory of Future Networks of Intelligence, The Chinese University of Hong Kong, Shenzhen, Guangdong 518172, China. (e-mail: xu.hui.zhang@foxmail.com; jinkeren@cuhk.edu.cn).
}

\thanks{Changsheng You is with the Department of Electronic and Electrical Engineering, Southern University of Science and Technology, Shenzhen 518055, China. (e-mails: youcs@sustech.edu.cn).
}

\thanks{Shuangyang Li is with the Department of Electrical Engineering and
 Computer Science, Technical University of Berlin, 10587 Berlin, Germany. (e-mails: shuangyang.li@tu-berlin.de).
}

\thanks{Corresponding author: Weijie Yuan}

}

\maketitle

\begin{abstract}
Unmanned aerial vehicle (UAV)-enabled integrated sensing and communication (ISAC) is regarded as a key enabler for next-generation wireless systems. However, conventional fixed-position antennas limit the ability of UAVs to fully exploit their inherent potential.
To overcome this limitation, we propose a UAV-enabled ISAC framework equipped with fluid antennas (FAs), where the mobility of antenna elements introduces additional spatial degrees of freedom to simultaneously enhance communication and sensing performance. A multi-objective optimization problem is formulated to maximize the communication rates of multiple users while minimizing the Cramér-Rao bound (CRB) for the angle estimation of a single target.
Due to excessively frequent updates of FA positions may lead to response delay, a three-timescale optimization framework is developed to jointly optimize transmit beamforming, FA positions, and UAV trajectory based on their characteristics.
To solve the non-convexity of the problem, an alternating optimization-based algorithm is developed to obtain a sub-optimal solution. Numerical results show that the proposed scheme significantly outperforms various benchmark schemes, validating the effectiveness of integrating the FA technology into the UAV-enabled ISAC systems. 

\end{abstract}

\section{Introduction}
\IEEEPARstart{T}{he} emerging low-altitude economy (LAE) represents a transformative paradigm that leverages airspace within a few hundred meters above the ground to support a wide range of applications \cite{10879807,10693833}. 
The foundation of its development and future expansion lies in the deployment of robust communication infrastructures, particularly low-altitude wireless networks (LAWNs) \cite{yuan2025ground}. Unlike conventional terrestrial networks, the LAWNs are specifically designed to meet the stringent requirements of aerial scenarios such as real-time monitoring, disaster response, and cargo delivery, offering ubiquitous connectivity with high reliability and low latency \cite{liu2025movable}. 
Unmanned aerial vehicles (UAVs), also referred to as drones or autonomous aerial vehicles (AAVs), are widely regarded as a key enabler for the LAWNs. Owing to their operational efficiency and cost-effectiveness, the UAVs play a central role in a variety of the LAE applications, thereby driving the rapid expansion of the LAE missions \cite{9468714,you2025hybrid}. However, the increasing scarcity of spectrum resources and the limited computation capabilities of the UAVs impose critical challenges on providing reliable and high-capacity communication services in the LAWNs, which in turn restrict sustainable growth of the LAE applications. 

Fortunately, fluid antennas (FAs), also known as movable antennas, have emerged as a promising solution to tackle these challenges \cite{wong2020fluid, zhu2024historical}.
Unlike conventional fixed-position antennas (FPAs), the FAs dynamically adapt their physical positions in response to the time-varying dynamics of wireless channels \cite{10286328, zhang2025movable}, which allows the system to opportunistically exploit the most favorable channel conditions in real time, thereby significantly improving signal reception and transmission efficiency in highly dynamic and congested environments \cite{9650760}. 
One of the most notable advantages of the FAs lies in their ability to mitigate interference and combat severe fading \cite{new2025ATutorial}, both of which are particularly prevalent in the UAV communications due to high mobility and frequent line-of-sight (LoS) blockages. 
By adjusting the FA positions, the UAV equipped with FAs can maintain robust connectivity and achieve superior throughput and latency performance, thus supporting mission-critical LAE applications \cite{10654366}. 

However, as the LAE missions over the LAWNs become increasingly complex, there is a growing demand for highly accurate and real-time environmental sensing. Conventional approaches, which rely on communication-dedicated systems or separate sensing devices, often face challenges in providing sufficient sensing capabilities without incurring additional energy consumption or spectrum overhead \cite{9737357}. Integrated sensing and communication (ISAC) has emerged as a promising solution, offering a unified framework that simultaneously supports communication services and environmental sensing \cite{10418473,cong2024near,you2025next}. By leveraging shared spectral resources and hardware platforms, ISAC enables UAVs to perform tasks such as target detection, trajectory tracking, obstacle avoidance, and remote sensing while maintaining reliable communication links \cite{9456851,11134125,11095312}. The dual-functional operation allows the UAVs to communicate and sense concurrently within the same hardware, significantly improving operational efficiency and safety, without further taxing limited resources such as spectrum, energy, or payload capacity. Such capabilities are particularly valuable in crowded or unpredictable low-altitude airspace, where the LAWNs must support highly dynamic LAE missions. 

Motivated by the growing demand for dual-functional ISAC applications in the LAWNs and the unique advantages of the FA technology, we propose in this paper a novel UAV-enabled ISAC system with FAs. 
By exploiting the dual mobility of the UAV trajectory and the FA positions, the proposed system establishes a highly adaptive ISAC framework. 
Since excessively frequent updates of FA positions may cause response delay, a three-timescale optimization framework is developed to jointly design transmit beamforming, FA positions, and UAV trajectory in accordance with their characteristics. 
Compared to the existing FPA-based ISAC systems, the FA-aided UAV-ISAC system involves dual-scale spatial optimization, i.e., the small-scale antenna position adjustment within the FA arrays to exploit local spatial channel variations, and the large-scale UAV trajectory optimization for achieving favorable propagation conditions across the service region. 
The key contributions of this paper are summarized as follows:
\begin{itemize}
    \item We study a UAV-enabled ISAC system, where a single UAV equipped with transmit and receive FAs simultaneously serves multiple communication users and senses a single target via uplink echo signals. A multi-objective optimization problem is formulated to jointly maximize the multiuser communication rate and minimize the Cramér-Rao bound (CRB) for target angle estimation.
    \item To reduce the response delay caused by excessively frequent FA position updates, a three-timescale optimization framework is proposed to jointly optimize transmit beamforming, FA positions, and UAV trajectory according to their characteristics.
    \item To address the non-convexity of the formulated problem, it is decomposed into four tractable subproblems. An efficient alternating optimization (AO)-based algorithm is developed to iteratively solve these subproblems, and its convergence properties and computational complexity are rigorously analyzed.
    \item Extensive simulations are conducted to validate the proposed framework, demonstrating that the FA-aided UAV-enabled ISAC system achieves significant performance gains over benchmark schemes and confirming the effectiveness of FA integration in supporting LAE missions over LAWNs.
\end{itemize}

\textit{Organizations:}
The remainder of this paper is organized as follows. Section II reviews the related works. Section III presents the UAV-enabled ISAC system incorporating the FA technology and formulates the corresponding multi-objective optimization problem. An AO-based algorithm is developed in Section IV, along with analyses of its computational complexity and convergence behavior. Numerical results are provided in Section V, and Section VI concludes the paper with summary remarks.

\section{Related Works}
The rapid development of UAV-enabled ISAC systems has created new opportunities for flexible wireless system design. Meanwhile, FA systems (FASs) have attracted growing attention due to their ability to dynamically adjust antenna positions within a confined region, thereby providing additional spatial degrees of freedom (DoFs). Although these two research directions have been studied individually, their integration remains largely unexplored. This section therefore reviews the most relevant studies on FASs and UAV-enabled ISAC systems.
\subsection{Fluid Antenna Systems}
The FAS has recently gained considerable attention due to its capability to dynamically adjust antenna positions within a feasible region to select the optimal signal port. The reconfigurability enables remarkable diversity gain and resilience against fading, even in highly spatially correlated environments \cite{new2025ATutorial, wong2020fluid}. Motivated by these advantages, several FAS-enabled frameworks have been proposed. For instance, a multiple access scheme exploiting the spatial reconfigurability of the FAS to achieve substantial performance improvements was proposed in \cite{new2024fluid}, while \cite{tang2025capacity} explored an uplink system where both the base station and user terminals were equipped with the FAs to maximize channel capacity. Moreover, the FAS has been integrated into the ISAC systems. In particular, \cite{ye2025SCNR} developed an FAS-assisted multiple-input and multiple-output (MIMO) ISAC framework to simultaneously enhance the communication quality and the radar sensing, and a two-dimensional FAS-aided ISAC system was developed in \cite{Wang2024Fluid} for multiuser MIMO downlink networks that jointly optimized the FA positions and the precoding. In addition to these core studies, the integration of the FAS with other emerging technologies has also been explored. For example, the UAV relay-enabled systems incorporating the FAS with non-orthogonal multiple access (NOMA) were investigated in \cite{abdou2024Sum-Rate}, and \cite{Ghadi2025UAV-Relay} examined the UAV-enabled multiuser communication where ground users were equipped with the FAS under a rate-splitting multiple access (RSMA) framework. Overall, these studies highlighted the broad applicability of the FAS in advancing the ISAC systems and further revealed its strong potential for enhancing the UAV-enabled communication paradigms.

\subsection{UAV-enabled ISAC Systems}
The UAV-enabled ISAC systems can simultaneously support communication and sensing tasks on aerial space, enabling efficient reuse of spectrum and hardware resources and thereby improving both spectral and energy efficiency \cite{9737357, 10233771}. By exploiting the controllable mobility of the UAVs, these systems can dynamically optimize communication links and task scheduling based on real-time sensing information, achieving superior coverage and adaptability in complex environments \cite{Meng2024UAV-Enabled}. Several representative studies have explored the UAV-enabled ISAC systems. In \cite{lyu2022joint}, a UAV-mounted dual-functional access point equipped with a vertical uniform linear array (ULA) was designed to simultaneously support multiuser communications and ground target sensing. In \cite{deng2023beamforming}, an adaptable ISAC framework was proposed, where sensing was performed on demand with configurable duration independent of communication time to improve efficiency. In \cite{Abdissa2024Joint}, the UAVs with edge caching capabilities were utilized to deliver cached content while performing target sensing. In \cite{Zhuang2025UAV-Enabled}, a full-duplex UAV-enabled ISAC framework was developed that integrated sensing with edge learning services. Furthermore, \cite{Jiang2024UAV-Enabled} proposed a UAV-enabled ISAC tracking framework that jointly estimated target position and velocity to enhance mobility-aware operations. Collectively, these studies demonstrated the potential of the UAV-enabled ISAC systems to efficiently integrate communication and sensing functionalities while providing flexible deployment and adaptive performance in dynamic wireless environments.

\section{System Model and Problem Formulation}
\begin{figure}[t]
	\centering
	\includegraphics[width=\linewidth]{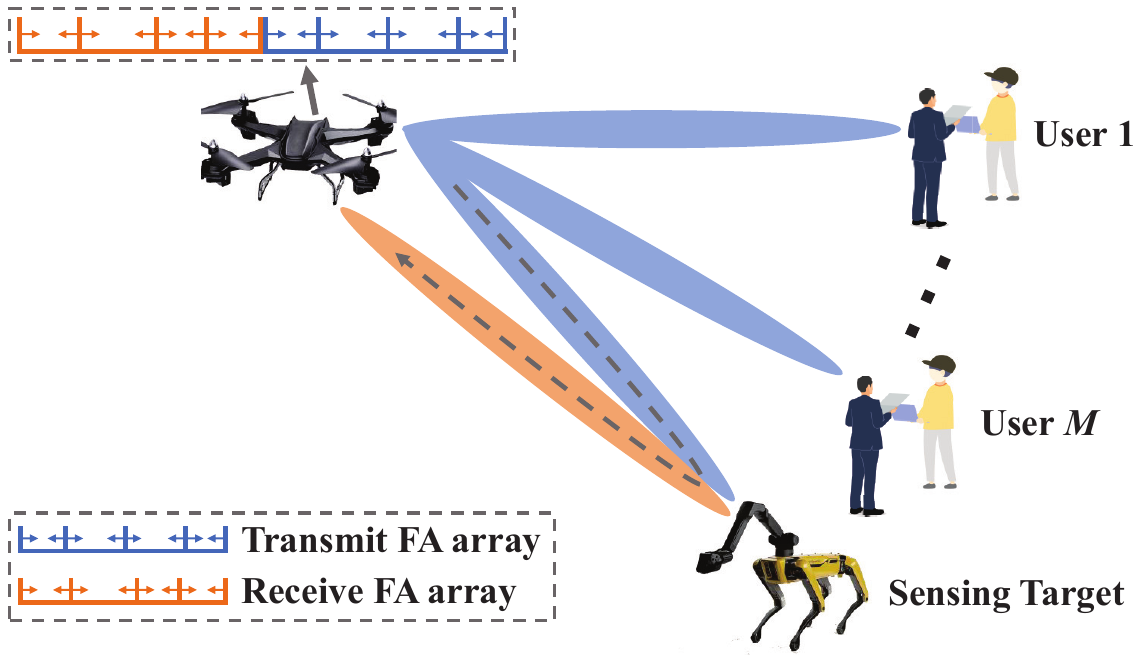}
	\caption{Illustration of the UAV-enabled ISAC system with transmit and receive FAs vertically mounted on the UAV.}
	\label{system model}
\end{figure}

As shown in Fig. ~\ref{system model}, a dual-function radar-communication UAV is deployed to serve $M$ downlink users and detect a single-point target. Let $\mathcal{M} = \{1, \dots, M\}$ denote the set of users. Specifically, the UAV transmits ISAC signals to provide communication services to users while simultaneously leveraging the reflected echoes from the target for radar sensing. Assuming that each user is equipped with a single antenna, and the UAV is equipped with two FA arrays of sizes $N_{t}$ and $N_{r}$ at the transmitter and receiver sides, respectively. 

Without loss of generality, a three-dimensional (3D) Cartesian coordinate system is considered. The UAV is assumed to fly at a fixed altitude $H$, starting from a given initial location $\mathbf{q}_{\rm{I}}$ and ending at a final location $\mathbf{q}_{\rm{F}}$. For analytical tractability, the UAV's flight duration is discretized into $N$ equal time slots, each of duration $\tau = T/N$, where $T$ denotes the total mission period. Let $\mathcal{N} = \{1, \dots, N\}$ denote the set of time slots.
The horizontal location of the UAV in time slot $n$ is denoted by $\mathbf{q}[n] = (x[n], y[n])$, and its full 3D location is therefore $(\mathbf{q}[n], H)$. The users are assumed to be static, with user $m$ located at $(\mathbf{q}_{m}, 0 )$, where $\mathbf{q}_{m} = (x_{m}, y_{m})$ denotes its horizontal location. Similarly, the target is located at $(\mathbf{q}_{\rm{T}}, 0)$, where $\mathbf{q}_{\rm{T}} = (x_{\rm{T}}, y_{\rm{T}})$ denotes its horizontal location.
Furthermore, the transmit and receive FA positions in time slot $n$ are defined as $\mathbf{x}[n] = \{ x_{1}[n], x_{2}[n], \ldots, x_{N_{t}}[n] \}^{\mathsf{T}}$ and $\mathbf{y}[n] = \{ y_{1}[n], y_{2}[n], \ldots, y_{N_{r}}[n]  \}^{\mathsf{T}}$, respectively~\footnote{
Due to the equivalence of the antennas, we can assume $x_{1}[n] < x_{2}[n] < \dots < x_{N_{t}}[n], \forall n \in \mathcal{N}$ and $y_{1}[n] < y_{2}[n] < \ldots < y_{N_{r}}[n], \forall n \in \mathcal{N}$.}.

\subsection{Channel Model}
We consider that the channel vector is determined not only by the signal propagation environment but also by the FA positions.
The far-field channel model is adopted, where the size of the transmit region is considerably smaller than the signal propagation distance \cite{abdou2024Sum-Rate, Ghadi2025UAV-Relay}.
Accordingly, all transmit FAs experience the same angle of departure (AoD) and the same amplitude of the complex path coefficient, while only the phase varies with their positions. 

Let $\theta_{m}[n]$ denote the vertical AoD from the UAV to the user $m$ in time slot $n$, which is given by
\begin{equation}
    \theta_{m}[n] = \arcsin \left( \frac{H}{\sqrt{\|\mathbf{q}[n] - \mathbf{q}_{m} \|^{2}+H^2}} \right). \label{theta_m}
\end{equation}
Then, the signal propagation phase difference between the $k$-th transmit FA position ($k = 1,2,\dots,N_{t}$) and the original position in time slot $n$ is given by $\frac{2\pi}{\lambda} x_{k}[n] \sin (\theta_{m}[n]) $, where $\lambda$ denotes the carrier wavelength.
Therefore, the transmit array response vector from the UAV to the user $m$ is given by  
\begin{equation}
\overline{\mathbf{h}}_{m}[n] = \left[
{\mathrm{e}}^{{\mathrm{j}} \frac{2 \pi}{\lambda} x_{1}[n] \sin (\theta_{m}[n]) },  \ldots, 
{\mathrm{e}}^{{\mathrm{j}} \frac{2 \pi}{\lambda} x_{N_{t}}[n] \sin (\theta_{m}[n]) }
\right]^{\mathsf{T}}. \label{h_los}
\end{equation}

In practical scenarios, the communication links between the UAV and users not only include a strong LoS component but may also involve a certain degree of multipath scattering. Therefore, a more general Rician fading model is adopted in this work. The channel vector from the UAV to the user $m$ in time slot $n$ is given by
\begin{equation} 
\mathbf{h}_{m}[n] = \sqrt{ \frac{ \kappa_{m}[n] \beta_{m}[n] }{ \kappa_{m}[n] + 1 } } \mathbf{\overline{h}}_{m}[n] + \sqrt{ \frac{ \beta_{m}[n] }{ \kappa_{m}[n] + 1 } } \mathbf{\tilde{h}}_{m}[n] ,
\label{comm_vec}
\end{equation}
where $\tilde{\mathbf{h}}_{m}[n]$ denotes the scattered Non-LoS (NLoS) component and is assumed to follow independent and identically distributed complex Gaussian distribution, i.e., $\tilde{\mathbf{h}}_{m}[n] \sim \mathcal{CN}(0, \mathbf{I}_{N_{t}})$.
$\beta_{m}[n] = h_0 (d_{m}[n])^{-2}$ represents the large-scale path-loss, where $h_0$ denotes the channel power gain at a reference distance of $1$ meter and $d_{m}[n] = \sqrt{\|\mathbf{q}[n] - \mathbf{q}_{m} \|^{2}+H^2}$ denotes the distance between the UAV and the user $m$.
Moreover, $\kappa_{m}[n] \ge 0$ denotes the Rician factor, characterizing the power ratio between the LoS and the NLoS components, which is expressed as\footnote{We consider the distance-dependent Rician factor, which can be directed expand to the system where the UAV can fly over the 3D airspace.}
\begin{equation}
\kappa_{m}[n] = c_{1} \exp{( c_{2} \theta_{m}[n]  )}, 
\end{equation}
where $c_{1}$ and $c_{2}$ are constant coefficients determined by the specific environment~\cite{you20193D}.

Next, the sensing channel model between the UAV and the target is considered. 
Following \cite{Wang2024Fluid}, a single-path propagation model is adopted in the considered ISAC system, as multipath reflections often suffer from significant attenuation in complex environments.
Therefore, the steering vectors of the transmit and receive FAs can be respectively given by
\begin{align}
& \mathbf{a}[n] = \left[
{\mathrm{e}}^{{\mathrm{j}} \frac{2 \pi}{\lambda} x_{1}[n] \sin (\theta_{\rm{T}}[n]) },  \ldots, 
{\mathrm{e}}^{{\mathrm{j}} \frac{2 \pi}{\lambda} x_{N_{t}}[n] \sin (\theta_{\rm{T}}[n]) }
\right]^{\mathsf{T}}, \label{transmit FA vector} \\ 
& \mathbf{b}[n] = \left[
{\mathrm{e}}^{{\mathrm{j}} \frac{2 \pi}{\lambda} y_{1}[n] \sin (\theta_{\rm{T}}[n]) },  \ldots, 
{\mathrm{e}}^{{\mathrm{j}} \frac{2 \pi}{\lambda} y_{N_{r}}[n] \sin (\theta_{\rm{T}}[n]) }
\right]^{\mathsf{T}}, \label{receive FA vector}
\end{align}
where $\theta_{\rm{T}}[n]$ denotes the vertical AoD from the UAV to the target in time slot $n$.

\subsection{User Communication Model}
We first consider the downlink communication from the UAV to the user $m$, where the UAV transmits an ISAC signal using coordinated transmit beamforming \footnote{In this paper, it is assumed that the UAV is equipped with a global positioning system (GPS) and can acquire the real-time location information of users through positioning systems. Accordingly, the large-scale fading is computed based on a geometric path-loss model, while the small-scale fading is characterized using distributions parameterized by the LoS probability. By combining these two components, the statistical channel state information (CSI) is generated for system design and performance analysis.}.
Let $\tilde{L} > N_{t}$ denote the length of the transmission frame in time slot $n$, where $\tilde{l} \in \{ 1,2,\dots, \tilde{L} \}$ represents the discrete time index within the frame. Specifically, the transmitted signal from the UAV in time slot $n$ is given by
\begin{equation}
   \mathbf{x}_{t}[\tilde{l},n] = \sum_{m=1}^{M} \mathbf{w}_{m}[n] s_{m}^{\mathrm{c}}[\tilde{l},n] + \mathbf{s}_{0}[\tilde{l},n], \label{transmitted signal}
\end{equation} 
where $\mathbf{w}_{m}[n] \in \mathbb{C}^{N_{t} \times 1}$ denotes the transmit beamforming vector directed toward the user $m$, and $s_{m}^{\mathrm{c}}[\tilde{l},n] \sim \mathcal{CN}(0,1)$ represents the corresponding information symbol, assumed to follow the circularly symmetric complex Gaussian (CSCG) distribution. $\mathbf{s}_{0}[\tilde{l},n]$ denotes the dedicated sensing signal, which follows a zero-mean CSCG distribution with covariance matrix $\mathbf{R}_{0}[n] = \mathbb{E} \{ \mathbf{s}_{0}[\tilde{l},n] \mathbf{s}^{\mathsf{H}}_{0}[\tilde{l},n] \} \succeq \mathbf{0}$.
\footnote{
It is noted that in \eqref{transmitted signal}, we consider a multi-beam transmission strategy for dedicated radar signals $\mathbf{s}_{0}[\tilde{l},n]$~\cite{lyu2022joint}. Such a design can be directly extended to the detection of multiple targets. Consequently, $\mathbf{R}_{0}[n]$ is of general rank, with $\text{rank}(\mathbf{R}_{0}[n]) = N_{\mathrm{s}}$, where $0 \leq N_{\mathrm{s}} \leq N_{t}$. This consideration corresponds to forming a set of $N_{\mathrm{s}}$ radar signal beams in time slot $n$, which can be obtained through the eigenvalue decomposition of $\mathbf{R}_{0}[n]$.}

Consequently, the signal received by the user $m$ is given by
\begin{align}
&z_{m}^{\mathrm{c}}[\tilde{l},n]  = \underbrace{ 
\mathbf{h}_{m}^{\mathsf{H}}[n]\mathbf{w}_{m}[n]s_{m}^{\mathrm{c}}[\tilde{l},n] }_{\rm{Desired \ signal}}  +
\underbrace{ \sum\limits_{ i \neq m }^{ \mathcal{M} } \mathbf{h}_{m}^{\mathsf{H}}[n] \mathbf{w}_{i}[n] s_{i}^{\mathrm{c}}[\tilde{l},n] }_{\rm{Inter-user\ interference}} \nonumber \\
& \quad\quad\quad\quad\  + \underbrace{ \mathbf{h}_{m}^{\mathsf{H}}[n] \mathbf{s}_{0}[\tilde{l},n]  }_{\rm{Sensing\ interference}}  + 
\underbrace{ \nu_{m}[\tilde{l},n] }_{\rm{AWGN}},     
\end{align}
where $\nu_{m}[\tilde{l},n] \sim \mathcal{CN}(0,\sigma^2_{m})$ is the additive white Gaussian noise (AWGN) at user $m$.
Then, the signal-to-interference-plus-noise ratio of the user $m$ in time slot $n$ is given by
\begin{equation} 
\gamma_{m}[n] = \frac{ \vert \mathbf{h}_{m}^{\mathsf{H}}[n]
 \mathbf{w}_{m}[n]  \vert^{2}  }
{ \sum\limits_{ i \neq m }^{ \mathcal{M} } 
\vert  \mathbf{h}_{m}^{\mathsf{H}}[n] \mathbf{w}_{i}[n] \vert^{2} 
 +  \mathbf{h}_{m}^{\mathsf{H}}[n] \mathbf{R}_{0}[n] \mathbf{h}_{m}[n] + \sigma^2_{m}}.
\end{equation}

\subsection{Target Sensing Model}
Next, we focus on the UAV's radar-based sensing of a single-point target.
Following~\cite{lyu2022joint}, it is assumed that the UAV has perfect knowledge of the communication signals. This allows the UAV to exploit the reflected communication signals for target sensing, which are therefore not treated as interference at the radar receiver.
To evaluate the sensing performance, the CRB is adopted as a fundamental performance metric~\cite{deng2023beamforming}, which provides a lower bound on the variance of any unbiased estimator. 

The target response matrix can be given by
\begin{equation}
    \mathbf{G}[n] = \frac{\varsigma_{\rm{T}}}{2d_{\rm{T}}[n]} \mathbf{b}[n] \mathbf{a}^{\mathsf{H}}[n], 
\end{equation}
where $d_{\rm{T}}[n]$ denotes the distance between the UAV and the target, and $\varsigma_{\rm{T}}$ denotes the complex radar cross-section
(RCS)~\cite{yuan2021Integrated}. Therefore, the received echo signal at the UAV for target sensing is given by
\begin{equation}
    \mathbf{Y}_{r}[n] = \mathbf{G}[n] \mathbf{X}_{t}[n]  + \mathbf{N}_{r}[n],
\end{equation}
where $\mathbf{X}_{t}[n] = \big[ \mathbf{x}_{t}[1,n],\mathbf{x}_{t}[2,n],\ldots,\mathbf{x}_{t}[\tilde{L},n] \big] \in \mathbb{C}^{N_{t} \times \tilde{L}}$, and $\mathbf{N}_{r}[n] \in \mathbb{C}^{N_{r} \times \tilde{L}}$ denotes an AWGN matrix with zero mean and the variance of each its entry is $\sigma_{r}^{2}$. Assuming that $\tilde{L}$ is sufficiently large, the covariance matrix of the transmitted signal can be approximated as
\begin{equation}
  \mathbf{R}_{\mathbf{x}}[n] = \frac{1}{\tilde{L}} \mathbf{X}_t[n] \mathbf{X}_t^{\mathsf{H}}[n] \approx \sum_{m=1}^{M} \mathbf{w}_m[n] \mathbf{w}_m^{\mathsf{H}}[n] + \mathbf{R}_{0}[n]. \label{covariance_Rx}
\end{equation}

According to~\cite{liu2022Cramér-Rao}, the CRB for estimating the elevation angle  $\theta_{\rm{T}}[n]$ of the target is given by \eqref{CRB_1}, where $\mathbf{A}[n] = \mathbf{b}[n] \mathbf{a}^{\mathsf{H}}[n]$ and $\dot{\mathbf{A}}[n] = \frac{ \partial \mathbf{A}[n]}{ \partial \theta_{\rm{T}}[n] }$. Due to the special structure of the $\text{CRB}(\theta_{\rm{T}}[n])$, we can reformulate it as \eqref{CRB_2}, and the detailed proof is provided in Appendix \ref{proof1}.

\begin{figure*}[!t]
\vspace*{-\baselineskip} 
{\small \begin{align} 
\text{CRB}(\theta_{\rm{T}}[n]) &= \frac{ \sigma_{r}^{2} }
{ 2 \left\vert \frac{\varsigma_{\rm{T}}}{2d_{\rm{T}}[n]} \right\vert^2 \tilde{L}
\left( \mathsf{tr}\left( \dot{\mathbf{A}}^{\mathsf{H}}[n] \dot{\mathbf{A}}[n] \mathbf{R}_{\mathbf{x}}[n] \right) - 
\frac{
\left\vert \mathsf{tr}\left( \dot{\mathbf{A}}^{\mathsf{H}}[n] \mathbf{A}[n] \mathbf{R}_{\mathbf{x}}[n] \right) \right\vert^2  }{
\mathsf{tr}\left( \mathbf{A}^{\mathsf{H}}[n] \mathbf{A}[n] \mathbf{R}_{\mathbf{x}}[n] \right)
}   \right)  } \nonumber \\
&= \frac{ 2 d_{\rm{T}}^{2}[n] \sigma_{r}^{2} \mathsf{tr}\left( \mathbf{A}^{\mathsf{H}}[n] \mathbf{A}[n] \mathbf{R}_{\mathbf{x}}[n] \right)}
{ \left\vert \varsigma_{\rm{T}} \right\vert^2 \tilde{L} \left( \mathsf{tr}\left( \dot{\mathbf{A}}^{\mathsf{H}}[n] \dot{\mathbf{A}}[n] \mathbf{R}_{\mathbf{x}}[n] \right) \mathsf{tr}\left( \mathbf{A}^{\mathsf{H}}[n] \mathbf{A}[n] \mathbf{R}_{\mathbf{x}}[n] \right) - \left\vert \mathsf{tr}\left( \dot{\mathbf{A}}^{\mathsf{H}}[n] \mathbf{A}[n] \mathbf{R}_{\mathbf{x}}[n] \right) \right\vert^2 \right)}.
\label{CRB_1}
\end{align} 
\vspace{-\baselineskip}
\begin{align}
\text{CRB}(\theta_{\rm{T}}[n]) = \frac{ 2 d_{\rm{T}}^{2}[n] \sigma_{r}^{2} }
 {  \vert \varsigma_{\rm{T}} \vert^2 \tilde{L} \left(\frac{2 \pi}{\lambda} \cos (\theta_{\rm{T}}[n]) \right)^{2} \mathbf{a}^{\mathsf{H}}[n] \mathbf{R}_{\mathbf{x}}[n] \mathbf{a}[n] \mathbf{y}[n]^{\mathsf{T}} \left( \mathbf{I}_{N_{r}} - \frac{1}{N_{r}} \mathbf{1}_{N_{r}} \mathbf{1}_{N_{r}}^{\mathsf{T}} \right) \mathbf{y}[n] }.  
 \label{CRB_2}
\end{align}
} \hrulefill
\end{figure*}

\subsection{Performance Metrics}
\textit{\textbf{Communication Metric}}: To ensure the long-term performance of the communication system, we adopt the ergodic sum rate of all users as the performance metric, which is mathematically given by
\begin{equation}
\mathbb{E} \left[ \sum_{n=1}^{N} \sum_{m=1}^{M} R_{m}[n] \right]
 = \mathbb{E} \left[ \sum_{n=1}^{N} \sum_{m=1}^{M} \log_{2} \left( 1 + \gamma_{m}[n]\right) \right]. 
\end{equation}

\textit{\textbf{Sensing Metric}}: To evaluate the target sensing performance, we adopt the CRB for the angle estimation of $\theta_{\rm{T}}[n]$ as the sensing performance metric. Based on the derivation in \eqref{CRB_2}, minimizing the CRB of the angle estimation is equivalent to maximizing the inverse of the CRB, i.e.,
\begin{align}
&\sum_{n=1}^{N} \frac{1}{\text{CRB}(\theta_{\rm{T}}[n])} \nonumber \\
& = \sum_{n=1}^{N} \alpha[n] \mathbf{a}^{\mathsf{H}}[n] \mathbf{R}_{\mathbf{x}}[n] \mathbf{a}[n] \mathbf{y}[n]^{\mathsf{T}} \left( \mathbf{I}_{N_{r}} - \frac{1}{N_{r}} \mathbf{1}_{N_{r}} \mathbf{1}_{N_{r}}^{\mathsf{T}} \right) \mathbf{y}[n] , \label{1/CRB}
\end{align}
where $\alpha[n] = \frac{ \vert \varsigma_{\rm{T}} \vert^2 \tilde{L} \left(\frac{2 \pi}{\lambda} \cos (\theta_{\rm{T}}[n]) \right)^{2} }
 { 2 d_{\rm{T}}^{2}[n] \sigma_{r}^{2} }$.

\subsection{Three-Timescale Optimization Framework}
To balance performance gains and practical implementation constraints, we adopt a three-timescale optimization framework. Although frequently updating the FA positions potentially enhance system performance, excessively frequent adjustments may incur additional control latency, increase energy consumption, and introduce non-ideal effects such as Doppler shifts and phase instability.

To address these issues, the system design is divided into three hierarchical timescales, as illustrated in Fig. 2. Specifically, the UAV trajectory is optimized at a long timescale over the entire mission period to account for large-scale propagation characteristics. The FA positions are updated at a medium timescale, i.e., in every $\mu$ time slots, to balance spatial flexibility and mechanical overhead. Meanwhile, the transmit beamforming is optimized at a short timescale, i.e., in every time slot, to adapt to instantaneous channel variations. In Fig. 2, $ \tilde{\mathcal{I}} = \{1, \dots, \tilde{I} \}$ denotes the set of intervals, and $N = \mu \tilde{I}$ with $\mu \in \mathbb{Z}$.

\begin{figure}[t]
	\centering
	\includegraphics[width=0.9\linewidth]{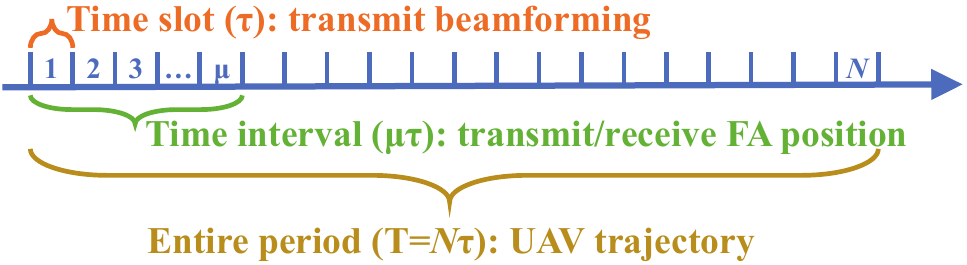}
	\caption{Illustration of three-timescale optimization framework.}
	\label{three-timescale}
\end{figure}

\subsection{Problem Formulation}
In this work, we aim to develop a joint optimization framework that balances the trade-off between two competing objectives: minimizing the CRB on the angle estimation (or equivalently, maximizing the inverse of the CRB, as defined in \eqref{1/CRB}) and maximizing the ergodic sum rate of all users. To achieve this, we introduce weight factors $\xi_{c}$ and $\xi_{s}$, representing the relative priorities of communication performance and sensing accuracy, respectively. 
Accordingly, the objective function of the joint optimization problem is given by
\begin{equation}
\mathcal{F} = \xi_{c} \mathbb{E} \left[ \sum_{n=1}^{N} \sum_{m=1}^{M} R_{m}[n] \right] + \xi_{s} \sum_{n=1}^{N} \frac{1}{\text{CRB}(\theta_{\rm{T}}[n])}.
\end{equation}

The objective is achieved by jointly optimizing the following variables: the UAV trajectory $\{\mathbf{q}[n]\}$, the beamforming design $\{ \mathbf{w}_{m}[n], \mathbf{R}_{0}[n] \}$, the transmit FA positions $ \{\mathbf{x}[n] \}$, and the receive FA positions $ \{\mathbf{y}[n] \}$.
Therefore, the optimization problem can be formulated as
\begin{subequations} 
\begin{flalign}
 (\textbf{P1}):\ & \max_{ \mathbf{q}[n],\mathbf{w}_{m}[n],\mathbf{R}_{0}[n],
 \mathbf{x}[n],\mathbf{y}[n] } \quad \mathcal{F} \label{p1a}  \\
 {\rm{s.t.}}  \quad &\mathbf{q}[1]= \mathbf{q}_{\rm{I}}, \quad
\mathbf{q}[N]=\mathbf{q}_{\rm{F}}, \label{p1b} \\
 & \| \mathbf{v}[n] \|  \le V_{\rm{max}}, \forall n\in \mathcal{N}\backslash\{1\}, \label{p1c} \\
 & \sum_{m= 1}^{M} \Vert \mathbf{w}_{m}[n] \Vert^2 + \mathsf{tr}(\mathbf{R}_{0}[n]) \le P_{\max},\ \forall n \in \mathcal{N}, \label{p1d}  \\
& \mathbf{x}[j] = \mathbf{x}[i\mu], \forall j \in \{(i-1)\mu+1,\dots,i\mu\}, i\in \tilde{\mathcal{I}}, \label{p1e} \\
 & \mathbf{y}[j] = \mathbf{y}[i\mu], \forall j \in \{(i-1)\mu+1,\dots,i\mu\}, i\in \tilde{\mathcal{I}}, \label{p1f} \\
 & \mathbf{P} \mathbf{x}[n] \succeq \mathbf{l}_{\rm{p}} , \forall n \in \mathcal{N}, \label{p1g} \\
 & \mathbf{Q} \mathbf{y}[n] \succeq \mathbf{l}_{\rm{q}} , \forall n \in \mathcal{N}, \label{p1h} 
\end{flalign} 
\end{subequations}
where constraints (\ref{p1b}) and (\ref{p1c}) describe the kinematic limitations of the UAV. Specifically, the UAV must depart from and return to the predetermined initial and final locations, denoted by $\mathbf{q}_{\rm{I}}$ and $\mathbf{q}_{\rm{F}}$, respectively. Additionally, its velocity in each time slot is constrained by the maximum allowable value $V_{\rm{max}}$. 
Constraint (\ref{p1d}) indicates that the transmit power of the UAV must not exceed a predefined limit $P_{\max}$. 
Constraints (\ref{p1e}) and (\ref{p1f}) ensure that the FA positions remain constant over the medium timescale.
Constraints (\ref{p1g}) and (\ref{p1h}) indicate that the distance between any two transmit/receive FAs must be greater than the minimum distance $ d_{\rm{min}} $ to avoid antenna coupling, and all antennas are required to move within a one-dimensional segment $[0, D_{\rm{FA}}]$. Specifically, the matrix $\mathbf{P} \in \mathbb{R}^{(N_{t} + 1) \times N_{t}}$ and vector $\mathbf{l}_{\rm{p}} \in \mathbb{R}^{(N_{t} + 1) \times 1}$ that define the constraints on the transmit FA positions are given by
\begin{equation}
\mathbf{P} = 
\begin{bmatrix}
-1 & 1 & 0 & 0 & \cdots & 0 & 0 \\
0 & -1 & 1 & 0 & \cdots & 0 & 0 \\
\vdots & \vdots & \vdots & \vdots & \ddots & \vdots & \vdots \\
0 & 0 & 0 & 0 & \cdots & -1 & 1 \\
1 & 0 & 0 & 0 & \cdots & 0 & 0 \\
0 & 0 & 0 & 0 & \cdots & 0 &  -1
\end{bmatrix}_{(N_{t}+1) \times N_{t}},
\end{equation}
\begin{equation}
\mathbf{l}_{\rm{p}} = [d_{\rm{min}}, d_{\rm{min}}, \cdots, d_{\rm{min}}, 0, -D_{\rm{FA}}]^{\mathsf{T}}.
\end{equation}
The matrices $\mathbf{Q}$ and $\mathbf{l}_{\rm{q}}$ that define the corresponding constraints for the receive FA positions have a similar structure and are omitted for brevity.

\section{Proposed Solution}
Due to the non-concavity of the objective function and the strong coupling among variables, problem (\textbf{P1}) is challenging to solve for a globally optimal solution. Furthermore, the UAV trajectory and the antenna positions are intricately embedded within the channel vector model. To address these, we first equivalently reformulate the problem by decomposing it into four subproblems. Subsequently, we employ an AO-based algorithm, iteratively optimizing the UAV trajectory, the beamforming design, the transmit FA positions, and the receive FA positions, in sequence, to obtain a sub-optimal solution.

\subsection{Transmit Beamforming Optimization} \label{beamforming_P}
Given the UAV trajectory $\{ \mathbf{q}[n] \}$ and the FA positions $ \{\mathbf{x}[n], \mathbf{y}[n] \}$, the optimization problem for the transmit beamforming is formulated as
\begin{subequations} 
\begin{flalign}
 (\textbf{P2}):\ & \max_{ \mathbf{w}_{m}[n],\mathbf{R}_{0}[n] } \quad \mathcal{F}_{1} \label{p3a}  \\
 {\rm{s.t.}}  \quad &\eqref{p1d}. \nonumber
\end{flalign}
\end{subequations}
Problem (\textbf{P2}) is challenging to solve directly due to the non-concavity of the objective function $\mathcal{F}$. To render the problem more tractable, we derive a ‌manageable approximation of the ergodic rate $\mathbb{E} \big[ R_{m}[n] \big]$ in $\mathcal{F}$ using the following proposition.
\begin{proposition}\label{proposition1}
The ergodic rate $\mathbb{E} \big[ R_{m}[n] \big]$ in the objective function $\mathcal{F}$ can be approximated as \eqref{ERate_1}, where $\zeta^{\rm{los}}_{m}[n] = \frac{ \kappa_{m}[n] \beta_{m}[n] }{ \kappa_{m}[n] + 1 }$ and $\zeta^{\rm{nlos}}_{m}[n] = \frac{ \beta_{m}[n] }{ \kappa_{m}[n] + 1 }$.
\begin{proof}
Please refer to Appendix \ref{proof2}.
\end{proof}
\end{proposition} 
\begin{figure*}[!t]
\vspace*{-\baselineskip} 
\begin{equation} \small
 \tilde{R}_{m}[n] = \log_{2} \left( 1 + \frac{ \zeta^{\rm{los}}_{m}[n] \left\vert \mathbf{\overline{h}}_{m}^{\mathsf{H}}[n]\mathbf{w}_{m}[n] \right\vert^{2} + \zeta^{\rm{nlos}}_{m}[n] \| \mathbf{w}_{m}[n] \|^{2} }
{ \sum\limits_{ i \neq m }^{ \mathcal{M} } 
\left( \zeta^{\rm{los}}_{m}[n] \left\vert \mathbf{\overline{h}}_{m}^{\mathsf{H}}[n]\mathbf{w}_{i}[n] \right\vert^{2} + \zeta^{\rm{nlos}}_{m}[n] \| \mathbf{w}_{i}[n] \|^{2} \right)
 + \zeta^{\rm{los}}_{m}[n] \mathbf{\overline{h}}_{m}^{\mathsf{H}}[n] \mathbf{R}_{0}[n] \mathbf{\overline{h}}_{m}[n] + \zeta^{\rm{nlos}}_{m}[n]\mathsf{tr}\left(\mathbf{R}_{0}[n] \right)   + \sigma^2_{m}} \right) .
 \label{ERate_1}
\end{equation}
\hrulefill
\end{figure*}

Therefore, the objective function $\mathcal{F}$ can be approximated as
\begin{equation}
\mathcal{F}_{1} = \xi_{c}  \sum_{n=1}^{N} \sum_{m=1}^{M} \tilde{R}_{m}[n]  + \xi_{s} \sum_{n=1}^{N} \frac{1}{\text{CRB}(\theta_{\rm{T}}[n])}, \label{p2g} 
\end{equation}
which is non-concave with respect to both $\mathbf{w}_{m}[n]$ and $\mathbf{R}_{0}[n]$. Furthermore, the optimization variables include both matrices and vectors, which further complicate the problem. To make the problem more tractable, we first define $ \mathbf{W}_{m}[n] =  \mathbf{w}_m[n] \mathbf{w}_m^{\mathsf{H}}[n]$, where $\mathbf{W}_{m}[n] \succeq \mathbf{0}$ and $\mathsf{rank}(\mathbf{W}_{m}[n]) = 1$. 

Therefore, $\tilde{R}_{m}[n]$ and $\frac{1}{\text{CRB}(\theta_{\rm{T}}[n])}$ can be rewritten as
\begin{equation}
\tilde{R}_{m}[n] = \log_{2} \left( 1 + \frac{E_{m}[n]}{F_{m}[n]} \right),
\end{equation}
\begin{align}
\frac{1}{\text{CRB}(\theta_{\rm{T}}[n])} = D_{m}[n] \mathbf{a}^{\mathsf{H}}[n] \mathbf{R}_{\mathbf{x}}[n] \mathbf{a}[n],
\end{align}
where
\begin{equation}
E_{m}[n] = \zeta^{\rm{los}}_{m}[n] \overline{\mathbf{h}}_{m}^{\mathsf{H}}[n] \mathbf{W}_{m}[n] \overline{\mathbf{h}}_{m}[n] + \zeta^{\rm{nlos}}_{m}[n] \mathsf{tr}\left( \mathbf{W}_{m}[n] \right),
\end{equation}

{\begin{align}
&F_{m}[n] 
= \zeta^{\rm{los}}_{m}[n] \overline{\mathbf{h}}_{m}^{\mathsf{H}}[n]
\mathbf{R}_{0}[n] \overline{\mathbf{h}}_{m}[n] + \zeta^{\rm{nlos}}_{m}[n]\mathsf{tr}\left(\mathbf{R}_{0}[n] \right) + \sigma^2_{m} \nonumber \\
& +
\sum_{i \neq m}^{M} \left( \zeta^{\rm{los}}_{m}[n] \overline{\mathbf{h}}_{m}^{\mathsf{H}}[n] \mathbf{W}_{i}[n] \overline{\mathbf{h}}_{m}[n] + \zeta^{\rm{nlos}}_{m}[n] \mathsf{tr}\left( \mathbf{W}_{i}[n] \right) \right),
\end{align} }
\begin{equation}
D_{m}[n] = \alpha[n] \mathbf{y}[n]^{\mathsf{T}} \left( \mathbf{I}_{N_{r}} - \frac{1}{N_{r}} \mathbf{1}_{N_{r}} \mathbf{1}_{N_{r}}^{\mathsf{T}} \right) \mathbf{y}[n]. 
\end{equation}

Furthermore, we recognize that problem (\textbf{P2}) constitutes a standard weighted sum-rate maximization problem. To address it, we employ the fractional programming (FP) for problem reformulation \cite{shen2018fractional}. Consequently, problem (\textbf{P2}) can be equivalently transformed into 
\begin{subequations} 
\begin{flalign}
 (\textbf{P2-1}):\ & \max_{ \mathbf{W}_{m}[n],\mathbf{R}_{0}[n],\omega_{m}[n], \varpi_{m}[n] } \quad \tilde{\mathcal{F}}_{1} \label{p3b}  \\
 {\rm{s.t.}}  \quad &  \mathsf{rank}(\mathbf{W}_{m}[n]) = 1, \forall m \in \mathcal{M}, \forall n \in \mathcal{N},\label{p3c} \\
 & \mathbf{W}_{m}[n], \mathbf{R}[n] \succeq \mathbf{0}, \forall m \in \mathcal{M}, \forall n \in \mathcal{N}, \label{p3d} \\
 & \sum_{m= 1}^{M} \mathsf{tr}\left( \mathbf{W}_{m}[n] \right) + \mathsf{tr}(\mathbf{R}_{0}[n]) \le P_{\max},\ \forall n \in \mathcal{N},  \label{p3e}\end{flalign}\end{subequations}where\begin{equation}
\tilde{\mathcal{F}}_{1} = \xi_{\text{c}}  \sum_{n=1}^{N} \sum_{m=1}^{M} \tilde{R}^{\rm{III}}_{m}[n]  + \xi_{\text{s}} \sum_{n=1}^{N} \frac{1}{\text{CRB}(\theta_{\rm{T}}[n])}. \label{p3f}
\end{equation}

Specifically, $\tilde{R}^{\rm{III}}_{m}[n]$ is given by
\begin{align}
\tilde{R}^{\rm{III}}_{m}[n] &= \log_2 \left( 1 + \omega_{m}[n] \right) + 2\varpi_{m}[n]\sqrt{E_{m}[n](1+\omega_{m}[n])} \nonumber \\
&\quad - \omega_{m}[n] - \varpi_{m}^{2}[n](E_{m}[n]+F_{m}[n]), 
\end{align}
where $\omega_{m}[n] \ge 0$ and $\varpi_{m}[n] \ge 0$ are auxiliary variables introduced in the FP-based reformulation.

Notably, when variables $\mathbf{W}_{m}[n]$ and $\mathbf{R}_{0}[n]$ are fixed, the objective function $\tilde{\mathcal{F}}_{1}$ is concave and differentiable with respect to the auxiliary variables $\omega_{m}[n]$ and $\varpi_{m}[n]$. The optimal solutions can be derived by setting the first-order derivatives to zero, i.e.,
\begin{equation}
    \omega_{m}^{\rm{opt}}[n] = \frac{E_{m}[n]}{F_{m}[n]},\ \varpi_{m}^{\rm{opt}}[n] = \frac{\sqrt{E_{m}[n](1+\omega_{m}[n])}}{E_{m}[n]+F_{m}[n]}. \label{auxiliary_variable}
\end{equation}

However, given $\omega_{m}[n]$ and $\varpi_{m}[n]$, problem (\textbf{P2-1}) remains non-convex with respect to $\mathbf{W}_{m}[n]$ due to the rank constraint \eqref{p3c}. To resolve this, we apply semi-definite relaxation (SDR) by removing \eqref{p3c}, yielding the convex problem,
\begin{subequations} 
\begin{flalign}
 (\textbf{P2-2}):\ & \max_{ \mathbf{W}_{m}[n],\mathbf{R}_{0}[n],\omega_{m}[n], \varpi_{m}[n] } \quad \tilde{\mathcal{F}}_{1}   \\
 {\rm{s.t.}}  \quad & \eqref{p3d}\text{ and }\eqref{p3e}, \nonumber
\end{flalign}
\end{subequations}
which can be efficiently solved via the CVX toolbox. To gain further insights, we analyze the Karush-Kuhn-Tucker (KKT) conditions of problem (\textbf{P2-2}), which confirms that problem (\textbf{P2-2}) always exists a rank-one optimal solution $\{ \mathbf{W}_{m}^{\rm{opt}}[n] \}$. This implies that the solution to problem (\textbf{P2-2}) is also optimal for problem (\textbf{P2-1}) without loss of optimality. The complete proof is provided in Appendix \ref{proof3}.

\subsection{Transmit FA Position Optimization}
Based on Proposition~\ref{proposition1}, given the UAV trajectory $\{ \mathbf{q}[n] \}$, the transmit beamforming $\{ \mathbf{w}_{m}[n], \mathbf{R}_{0}[n] \}$, and the receive FA positions $ \{\mathbf{y}[n]\}$, the optimization problem for the transmit FA positions is formulated as
\begin{subequations} 
\begin{flalign}
 (\textbf{P3}):\ & \max_{ \mathbf{x}[n] } \quad \mathcal{F}_{1} \label{p4a}  \\
 {\rm{s.t.}}  \quad &\eqref{p1e}\text{ and }\eqref{p1g}. \nonumber
\end{flalign}
\end{subequations}
Problem (\textbf{P3}) is challenging to solve directly due to the non-concavity of the objective function $\mathcal{F}$.
Following an approach similar to that presented in Section~\ref{beamforming_P}, problem (\textbf{P3}) can be equivalently transformed into
\begin{subequations} 
\begin{flalign}
 (\textbf{P3-1}):\ & \max_{ \mathbf{x}[n],\omega_{m}[n], \varpi_{m}[n] } \quad \tilde{\mathcal{F}}_{1} \label{p4b}  \\
 {\rm{s.t.}}  \quad & \eqref{p1e}\text{ and }\eqref{p1g}, \nonumber
\end{flalign}
\end{subequations}
where the objective function $\tilde{\mathcal{F}}_{1}$ is defined in \eqref{p3f}.

Let $[ \mathbf{W}_{m}[n] ]_{p,q}$, $\vert 
[ \mathbf{W}_{m}[n]]_{p,q} \vert$, and $\angle \left( [\mathbf{W}_{m}[n]]_{p,q} \right)$ denote the entry, the magnitude, and the phase, respectively, at the $p$-th row and $q$-th column of $\mathbf{W}_{m}[n]$.  Additionally, define $\vartheta_{m}[n] =  \frac{2 \pi}{\lambda}\sin (\theta_{m}[n])$.
By examining the structure of $\overline{\mathbf{h}}_{m}^{\mathsf{H}}[n] \mathbf{W}_{m}[n] \overline{\mathbf{h}}_{m}[n]$, we can obtain
\begin{align}
&\overline{\mathbf{h}}_{m}^{\mathsf{H}}[n] \mathbf{W}_{m}[n] \overline{\mathbf{h}}_{m}[n] \nonumber \\
&= \sum_{p=1}^{N_{t}} \sum_{q=1}^{N_{t}} [ \mathbf{W}_{m}[n] ]_{p,q} {\mathrm{e}}^{{\mathrm{j}} \frac{2 \pi}{\lambda} (x_{q}[n] - x_{p}[n]) 
\sin (\theta_{\rm{m}}[n])} \nonumber \\
& = \sum_{p=1}^{N_{t}} \sum_{q=1}^{N_{t}} 
\vert [ \mathbf{W}_{m}[n]]_{p,q} \vert 
\cos \left( \Psi_{m,p,q}[n] \right),
\end{align} 
where $\Psi_{m,p,q}[n] =  \vartheta_{m}[n] (x_{p}[n] - x_{q}[n]) - \angle \left( [\mathbf{W}_{m}[n]]_{p,q} \right).$ Since $\overline{\mathsf{h}}_{m}^{\mathrm{H}}[n] \mathbf{W}_{m}[n] \overline{\mathbf{h}}_{m}[n]$ is non-concave with respect to $\mathbf{x}[n]$, we construct a lower bound surrogate function to locally approximate it using its second-order Taylor expansion~\cite{ma2024Multi-Beam}.
Specifically, for a given $a \in \mathbb{R}$, we can obtain
\begin{align}
\cos (a) &\approx \cos (a_0) - \sin (a_0) (a -a_0) - \frac{1}{2} \cos (a_0) (a -a_0)^{2} \nonumber \\
& \geq \cos (a_0) - \sin (a_0) (a -a_0) - \frac{1}{2} (a -a_0)^{2} \nonumber \\
&= f\left(a \mid a_0 \right).
\end{align}
Therefore, for given $\mathbf{x}^{(l)}[n] = \left\{ x_{1}^{(l)}[n], x_{2}^{(l)}[n], \ldots, x_{N_{t}}^{(l)}[n] \right\}^{\mathsf{T}}$ in the $i$-th iteration of SCA, by letting $a_0 = \Psi_{m,p,q}^{(l)}[n]$ and $ a = \Psi_{m,p,q}[n]$, we can obtain
\begin{align}
&\overline{\mathbf{h}}_{m}^{\mathsf{H}}[n] \mathbf{W}_{m}[n] \overline{\mathbf{h}}_{m}[n] \nonumber \\
&\geq \sum_{p=1}^{N_{t}} \sum_{q=1}^{N_{t}} 
\vert [ \mathbf{W}_{m}[n]]_{p,q} \vert 
f\left( \Psi_{m,p,q}[n] \mid \Psi_{m,p,q}^{(l)}[n] \right) \nonumber \\
& = \frac{1}{2} \mathbf{x}^{\mathsf{T}}[n] \mathbf{S}^{\rm{I}}_{m}[n] \mathbf{x}[n] +\left(\mathbf{t}_{m}^{\rm{I}}[n]\right)^{\mathsf{T}} \mathbf{x}[n] + u_{m}^{\rm{I}}[n],
\end{align}
where $\mathbf{S}^{\rm{I}}_{m}[n] \in \mathbb{R}^{N_{t} \times N_{t}}$, $\mathbf{t}^{\rm{I}}_{m}[n] \in \mathbb{R}^{N_{t} \times 1}$, and $u^{\rm{I}}_{m}[n] \in \mathbb{R}$ are defined in \eqref{p4c}-\eqref{p4d}. 
\begin{figure*}[!t]
\vspace*{-\baselineskip} 
{ \small
\begin{align}
\mathbf{S}^{\rm{I}}_{m}[n] = -2 \vartheta_{m}^{2}[n]
\left( \mathsf{diag} \left( \mathbf{V}_{m}^{\rm{I}}[n] \right)-\mathbf{V}_{m}^{\rm{II}}[n]\right), 
\mathbf{V}_{m}^{\rm{I}}[n] = \sum_{q=1}^{N_{t}} \big[ \vert [\mathbf{W}_{m}[n]]_{1,q} \vert, \ldots, \vert [ \mathbf{W}_{m}[n]]_{N_{t},q} \vert \big]^{\mathsf{T}},\
\mathbf{V}_{m}^{\rm{II}}[n] = \left( \vert [\mathbf{W}_{m}[n]]_{p,q} \vert \right)_{p,q},
\label{p4c}
\end{align}
\vspace{-\baselineskip}
\begin{align}
[\mathbf{t}^{\rm{I}}_{m}[n]]_{p} = \sum_{q=1}^{N_{t}} \vert [ \mathbf{W}_{m}[n]]_{p,q} \vert \left( 2 \vartheta_{m}^{2}[n] \left(x_{p}^{(l)}[n] - x_{q}^{(l)}[n]\right) -
 2\vartheta_{m}[n] \sin \left( \Psi^{(l)}_{m,p,q}[n] \right) \right),
\end{align}
\vspace{-\baselineskip}
\begin{align}
u^{\rm{I}}_{m}[n] = & \sum_{p=1}^{N_{t}} \sum_{q=1}^{N_{t}} \vert [ \mathbf{W}_{m}[n]]_{p,q} \vert \left( \cos \left( \Psi^{(l)}_{m,p,q}[n] \right) + \vartheta_{m}[n] \sin \left( \Psi^{(l)}_{m,p,q}[n] \right) \left(x_{p}^{(l)}[n] - x_{q}^{(l)}[n]\right)
- \frac{1}{2} \vartheta_{m}^{2}[n]
\left(x_{p}^{(l)}[n] - x_{q}^{(l)}[n]\right)^{2}  \right). \label{p4d}
\end{align}
} \hrulefill
\end{figure*}
By introducing a slack variable $\phi_{m}^{\rm{I}}[n]$, we obtain the standard quadratic constraint (QC), i.e.,
\begin{equation} 
\frac{1}{2} \mathbf{x}^{\mathsf{T}}[n] \mathbf{S}^{\rm{I}}_{m}[n] \mathbf{x}[n] +\left(\mathbf{t}_{m}^{\rm{I}}[n]\right)^{\mathsf{T}} \mathbf{x}[n] + u^{\rm{I}}_{m}[n] \geq \phi_{m}^{\rm{I}}[n]. \label{p4e}
\end{equation}
Since $\mathbf{S}^{\rm{I}}_{m}[n]$ is negative semi-definite, constraint \eqref{p4e} is convex with respect to $\mathbf{x}[n]$. The complete proof is provided in Appendix \ref{proof4}.

On the other hand, we can get an upper bound surrogate function to locally approximate it using its second-order Taylor expansion.
Specifically, for a given $a \in \mathbb{R}$, we can obtain
\begin{align}
\cos (a) &\approx \cos (a_0) - \sin (a_0) (a -a_0) - \frac{1}{2} \cos (a_0) (a -a_0)^{2} \nonumber \\
& \leq \cos (a_0) - \sin (a_0) (a -a_0) + \frac{1}{2} (a -a_0)^{2} \nonumber \\
&= \tilde{f}\left(a \mid a_0 \right).
\end{align}
Therefore, we can construct the upper bound, i.e.,
\begin{align}
&\overline{\mathbf{h}}_{m}^{\mathsf{H}}[n] \mathbf{W}_{m}[n] \overline{\mathbf{h}}_{m}[n] \nonumber \\
&\leq \sum_{p=1}^{N_{t}} \sum_{q=1}^{N_{t}} 
\vert [ \mathbf{W}_{m}[n]]_{p,q} \vert 
\tilde{f}\left( \Psi_{m,p,q}[n] \mid \Psi_{m,p,q}^{(l)}[n] \right) \nonumber \\
& = \frac{1}{2} \mathbf{x}^{\mathsf{T}}[n] \tilde{\mathbf{S}}^{\rm{I}}_{m}[n] \mathbf{x}[n] +\left(\tilde{\mathbf{t}}_{m}^{\rm{I}}[n]\right)^{\mathsf{T}} \mathbf{x}[n] + \tilde{u}^{\rm{I}}_{m}[n],
\end{align}
where $\tilde{\mathbf{S}}^{\rm{I}}_{m}[n] \in \mathbb{R}^{N_{t} \times N_{t}}$, $\tilde{\mathbf{t}}^{\rm{I}}_{m}[n] \in \mathbb{R}^{N_{t} \times 1}$, and $\tilde{u}^{\rm{I}}_{m}[n] \in \mathbb{R}$ are defined in \eqref{p4f}-\eqref{p4g}. 
\begin{figure*}[!t]
\vspace*{-\baselineskip} 
{ \small
\begin{align}
\tilde{\mathbf{S}}^{\rm{I}}_{m}[n] = 2 \vartheta_{m}^{2}[n]
\left( \mathsf{diag} \left( \mathbf{V}_{m}^{\rm{I}}[n] \right)-\mathbf{V}_{m}^{\rm{II}}[n]\right), \label{p4f}
\end{align}
\vspace{-\baselineskip}
\begin{align}
[\tilde{\mathbf{t}}^{\rm{I}}_{m}[n]]_{p} = \sum_{q=1}^{N_{t}} \vert [ \mathbf{W}_{m}[n]]_{p,q} \vert \left( -2 \vartheta_{m}^{2}[n] \left(x_{p}^{(l)}[n] - x_{q}^{(l)}[n]\right) -
 2\vartheta_{m}[n] \sin \left( \Psi^{(l)}_{m,p,q}[n] \right) \right),
\end{align}
\vspace{-\baselineskip}
\begin{align}
\tilde{u}^{\rm{I}}_{m}[n] = & \sum_{p=1}^{N_{t}} \sum_{q=1}^{N_{t}} \vert [ \mathbf{W}_{m}[n]]_{p,q} \vert \left( \cos \left( \Psi^{(l)}_{m,p,q}[n] \right) + \vartheta_{m}[n] \sin \left( \Psi^{(l)}_{m,p,q}[n] \right) \left(x_{p}^{(l)}[n] - x_{q}^{(l)}[n]\right)
+ \frac{1}{2} \vartheta_{m}^{2}[n]
\left(x_{p}^{(l)}[n] - x_{q}^{(l)}[n]\right)^{2}  \right). \label{p4g}
\end{align}
} \hrulefill
\end{figure*}
By introducing a slack variable $\psi_{m}^{\rm{I}}[n]$, we obtain the standard QC, i.e.,
\begin{equation} 
\frac{1}{2} \mathbf{x}^{\mathsf{T}}[n] \tilde{\mathbf{S}}^{\rm{I}}_{m}[n] \mathbf{x}[n] +\left(\tilde{\mathbf{t}}_{m}^{\rm{I}}[n]\right)^{\mathsf{T}} \mathbf{x}[n] + \tilde{u}^{\rm{I}}_{m}[n] \leq \psi_{m}^{\rm{I}}[n]. \label{p4h}
\end{equation}
Since $\mathbf{S}^{\rm{I}}_{m}[n]$ is positive semi-definite, constraint \eqref{p4h} is concave with respect to $\mathbf{x}[n]$.

Similarly, we can construct the lower bound and upper bound surrogate functions for $\mathbf{a}^{\mathsf{H}}[n] \mathbf{R}_{\mathbf{x}}[n] \mathbf{a}[n]$ in $\frac{1}{\text{CRB}(\theta_{\rm{T}}[n])}$, 
as well as for $\overline{\mathbf{h}}_{m}^{\mathsf{H}}[n] \mathbf{W}_{i}[n] \overline{\mathbf{h}}_{m}[n]$ and $\overline{\mathbf{h}}_{m}^{\mathsf{H}}[n] \mathbf{R}_{0}[n] \overline{\mathbf{h}}_{m}[n]$ in $F_{m}[n]$. Accordingly, we derive the associated constraints based on these surrogate functions, which take the following general forms,
\begin{equation} \small
\frac{1}{2} \mathbf{x}^{\mathsf{T}}[n] \mathbf{S}^{j}_{\chi}[n] \mathbf{x}[n] +\left(\mathbf{t}_{\chi}^{j}[n]\right)^{\mathsf{T}} \mathbf{x}[n] + u_{\chi}^{j}[n] \geq \phi_{\chi}^{j}[n],\chi \in \mathcal{M} \cup \{\rm{T}\}, \label{p4i}
\end{equation}
\begin{equation} \small
\frac{1}{2} \mathbf{x}^{\mathsf{T}}[n] \tilde{\mathbf{S}}^{j}_{\chi}[n] \mathbf{x}[n] +\left(\tilde{\mathbf{t}}_{\chi}^{j}[n]\right)^{\mathsf{T}} \mathbf{x}[n] + \tilde{u}^{j}_{\chi}[n] \leq \psi_{\chi}^{j}[n], \chi \in \mathcal{M} \cup \{\rm{T}\},  \label{p4j}
\end{equation}
where  the index $j = {{\rm{II}}(i)}, {\rm{III}},$ and ${\rm{IV}} $ corresponds to the terms $\overline{\mathbf{h}}_{m}^{\mathsf{H}}[n] \mathbf{W}_{i}[n] \overline{\mathbf{h}}_{m}[n]$, $\overline{\mathbf{h}}_{m}^{\mathsf{H}}[n] \mathbf{R}_{0}[n] \overline{\mathbf{h}}_{m}[n]$, and $\mathbf{a}^{\mathsf{H}}[n] \mathbf{R}_{\mathbf{x}}[n] \mathbf{a}[n]$, respectively.

Based on the above, the lower bound of objective function is given by
\begin{align}
\tilde{\mathcal{F}}^{\rm{lb}}_{1} = \xi_{c}  \sum_{n=1}^{N} \sum_{m=1}^{M} \tilde{R}^{\rm{lb}}_{m}[n]  + \xi_{s} \sum_{n=1}^{N} \frac{1}{\text{CRB}^{\rm{ub}}(\theta_{\rm{T}}[n])},
\end{align}
where 
\begin{align}
\tilde{R}^{\rm{lb}}_{m}[n] &= \log_2 \left( 1 + \omega_{m}[n] \right) + 2\varpi_{m}[n]\sqrt{E^{\rm{lb}}_{m}[n](1+\omega_{m}[n])} \nonumber \\
&\quad - \omega_{m}[n] - \varpi_{m}^{2}[n](E^{\rm{ub}}_{m}[n]+F_{m}^{\rm{ub}}[n]), 
\end{align}
\begin{equation}
E_{m}^{\rm{lb}}[n] = \zeta^{\rm{los}}_{m}[n] \phi_{m}^{\rm{I}}[n] + \zeta^{\rm{nlos}}_{m}[n] \mathsf{tr}\left( \mathbf{W}_{m}[n] \right),
\end{equation}
\begin{equation}
E_{m}^{\rm{ub}}[n] = \zeta^{\rm{los}}_{m}[n] \psi_{m}^{\rm{I}}[n] + \zeta^{\rm{nlos}}_{m}[n] \mathsf{tr}\left( \mathbf{W}_{m}[n] \right),
\end{equation}
 \begin{align}
F_{m}^{\rm{ub}}[n] &= \sum_{i = 1, i \neq m}^{M} \left( \zeta^{\rm{los}}_{m}[n] \psi_{m}^{{\rm{II}}(i)}[n] + \zeta^{\rm{nlos}}_{m}[n] \mathsf{tr}\left( \mathbf{W}_{i}[n] \right) \right) \nonumber \\
& \quad + \zeta^{\rm{los}}_{m}[n] \psi_{m}^{\rm{III}}[n] + \zeta^{\rm{nlos}}_{m}[n]\mathsf{tr}\left(\mathbf{R}_{0}[n] \right) + \sigma^2_{m},
\end{align} 
\begin{align}
\frac{1}{\text{CRB}^{\rm{ub}}(\theta_{\rm{T}}[n])} = D_{m}[n] \phi_{\rm{T}}^{\rm{IV}}[n].
\end{align}

Therefore, problem (\textbf{P3-1}) can be reformulated as
\begin{subequations} 
\begin{flalign}
 (\textbf{P3-2}):\ & \max_{ \mathbf{x}[n],\omega_{m}[n], \varpi_{m}[n] } \quad \tilde{\mathcal{F}}^{\rm{lb}}_{1} \label{p4k}  \\
 {\rm{s.t.}}  \quad & \eqref{p1e}, \eqref{p1g}, \eqref{p4e}, \eqref{p4h}, \eqref{p4i},\text{ and } \eqref{p4j}. \nonumber
\end{flalign}
\end{subequations}
Problem (\textbf{P3-2}) is a standard convex optimization problem, which can be efficiently solved using the CVX toolbox.

\subsection{Receive FA Position Optimization}
Obviously, the ergodic sum rate of all users is independent of the receive FA positions. Therefore, given the UAV trajectory $\{ \mathbf{q}[n] \}$, the transmit beamforming $\{ \mathbf{w}_{m}[n], \mathbf{R}_{0}[n] \}$, and the transmit FA positions $ \{\mathbf{x}[n]\}$, the optimization problem for the receive FA positions can be formulated as
\begin{subequations} 
\begin{flalign}
 (\textbf{P4}):\ & \max_{ \mathbf{y}[n] } \quad \sum_{n=1}^{N} \varphi[n] \mathbf{y}[n]^{\mathsf{T}} \left( \mathbf{I}_{N_{r}} - \frac{1}{N_r} \mathbf{1}_{N_{r}} \mathbf{1}_{N_{r}}^T \right) \mathbf{y}[n]  \label{p5a} \\
 {\rm{s.t.}}  \quad & \eqref{p1f}\text{ and }\eqref{p1h}, \nonumber
\end{flalign}
\end{subequations}
where $\varphi[n] = \xi_{\text{s}} \frac{ \vert \varsigma_{\rm{T}} \vert^2 \tilde{L} \left(\frac{2 \pi}{\lambda} \cos (\theta_{\rm{T}}[n]) \right)^{2} }
 { 2 d_{\rm{T}}^{2}[n] \sigma_{r}^{2} } \mathbf{a}^{\mathsf{H}}[n] \mathbf{R}_{\mathbf{x}}[n] \mathbf{a}[n]$.

In statistical theory, given a set of observed data $ \{ y_1, y_2, \cdots, y_n \} $, the sample mean and the total sum of squares (TSS) are defined as $\bar{y} = \frac{1}{n} \sum_{i=1}^{n} y_i$ and $\text{TSS} = \sum_{i=1}^{n} (y_i - \bar{y})^2$, respectively. Based on these definitions, the quadratic form $\mathbf{y}[n]^{\mathsf{T}} \left( \mathbf{I}_{N_{r}} - \frac{1}{N_{r}} \mathbf{1}_{N_{r}} \mathbf{1}_{N_{r}}^{\mathsf{T}} \right) \mathbf{y}[n]$ is equivalent to the TSS of the sequence $\mathbf{y}[n]$, and is equal to the sample variance scaled by a factor of $(N_{r} - 1)$. Therefore, the objective function corresponds to the weighted sum of the TSS across all time slots.

Therefore, the optimal solution to problem (\textbf{P4}) is given by the sequence $\{ y_{p}[n] \}_{p=1}^{N_{r}}$, expressed as
\begin{equation} \label{receive array solution}
y_{p}^{\rm{opt}}[n] = \begin{cases} 
(p-1)d_{\rm{min}}, \\ 
\quad\quad\quad\quad \text{if $N_{r}$ is even}, p = 1, \ldots, \frac{N_{r}}{2}, \\
D_{\rm{FA}} - (N_{r} - p)d_{\rm{min}}, \\ 
\quad\quad\quad\quad \text{if $N_{r}$ is even}, p = \frac{N_{r}}{2}+1, \ldots, N_{r}, \\
(p-1)d_{\rm{min}}, \\ 
\quad\quad\quad\quad \text{if $N_{r}$ is odd}, p = 1, \ldots, \frac{N_{r}-1}{2}, \\
D_{\rm{FA}} - (N_{r} - p)d_{\rm{min}}, \\ 
\quad\quad\quad\quad \text{if $N_{r}$ is odd}, p = \frac{N_{r}-1}{2} +1, \ldots, N_{r}.
\end{cases}
\end{equation}
This implies that, without increasing the overall array size, placing additional antennas at the boundaries of the array can enhance the sensing performance. The complete proof is provided in Appendix \ref{proof5}.

\subsection{UAV Trajectory Optimization}
Given the transmit beamforming $\{ \mathbf{w}_{m}[n], \mathbf{R}_{0}[n] \}$ and the FA positions $ \{\mathbf{x}[n], \mathbf{y}[n] \}$, the optimization problem for the UAV trajectory is formulated as
\begin{subequations}  
\begin{flalign}
 (\textbf{P5}):\ & \max_{ \mathbf{q}[n] } \quad \mathcal{F} \label{p2a} \\
 {\rm{s.t.}}  \quad &\eqref{p1b}\text{ and } \eqref{p1c}.\nonumber
\end{flalign}
\end{subequations}

As shown in \eqref{h_los} and \eqref{transmit FA vector}, the transmit steering vector $\mathbf{\overline{h}}_{m}[n]$ and $\mathbf{a}[n]$ are non-convex and non-linear functions of the UAV trajectory $\mathbf{q}[n]$, which makes the trajectory optimization particularly challenging.
To facilitate the UAV trajectory optimization, we approximate $\mathbf{\overline{h}}_{m}[n]$ and $\mathbf{a}[n]$ in the $(l+1)$-th iteration using the trajectory obtained in the $(l)$-th iteration. Accordingly, the approximated expressions for $\mathbf{\overline{h}}_{m}^{(l)}[n]$ and $\mathbf{a}^{(l)}[n]$ can be given by
\begin{equation} 
\overline{\mathbf{h}}_{m}^{(l)}[n] = \left[
{\mathrm{e}}^{{\mathrm{j}} \frac{2 \pi}{\lambda} x_{1}[n] \sin (\theta_{m}^{(l)}[n]) },  \ldots, 
{\mathrm{e}}^{{\mathrm{j}} \frac{2 \pi}{\lambda} x_{N_{t}}[n] \sin (\theta_{m}^{(l)}[n]) }
\right]^{\mathsf{T}},
\end{equation}
\begin{equation} 
\mathbf{a}^{(l)}[n] = \left[
{\mathrm{e}}^{{\mathrm{j}} \frac{2 \pi}{\lambda} x_{1}[n] \sin (\theta_{\rm{T}}^{(l)}[n]) },  \ldots, 
{\mathrm{e}}^{{\mathrm{j}} \frac{2 \pi}{\lambda} x_{N_{t}}[n] \sin (\theta_{\rm{T}}^{(l)}[n]) }
\right]^{\mathsf{T}}, 
\end{equation}
where $\theta_{i}^{(l)}[n] = \arcsin \left( \frac{H}{\sqrt{\|\mathbf{q}^{(l)}[n] - \mathbf{q}_{i} \|^{2}+H^2}} \right), i \in \mathcal{M} \cup \{\rm{T}\}$.
Therefore, the objective function $\mathcal{F}_{1}$ can be reformulated as
\begin{equation}
\mathcal{F}_{1}^{(l)} = \xi_{\text{c}}  \sum_{n=1}^{N} \sum_{m=1}^{M} \tilde{R}_{m}^{(l)}[n]  + \xi_{\text{s}} \sum_{n=1}^{N} \frac{1}{\text{CRB}^{(l)}(\theta_{\rm{T}}[n])},
\end{equation}
where 
\begin{equation}
\tilde{R}_{m}^{(l)}[n] = \log_{2} \left( 1 + \frac{ d_{m}^{-2}[n] B_{m}^{(l)}[n] }
{ d_{m}^{-2}[n] C_{m}^{(l)}[n] + \sigma^2_{m} } \right), \label{p2b}
\end{equation}
\begin{equation}
\frac{1}{\text{CRB}^{(l)}(\theta_{\rm{T}}[n])} = d_{\rm{T}}^{-2}[n] A^{(l)}[n], \label{p2c}
\end{equation}
with
\begin{equation}
B_{m}^{(l)}[n] = \frac{ \kappa_{m}^{(l)}[n] h_0 }{ \kappa_{m}^{(l)}[n] + 1 } \left| \left( \mathbf{\overline{h}}_{m}^{(l)}[n] \right)^{\mathsf{H}}  \mathbf{w}_{m}[n] \right|^{2} + \frac{ h_0 \| \mathbf{w}_{m}[n] \|^{2} }{ \kappa_{m}^{(l)}[n] + 1 } ,
\end{equation} 
\begin{align}
C_{m}^{(l)}[n] & = \sum\limits_{ i = 1, i \neq m }^{ \mathcal{M} } \left( \frac{ \kappa_{m}^{(l)}[n] h_0 }{ \kappa_{m}^{(l)}[n] + 1 } \left| \left( \mathbf{\overline{h}}_{m}^{(l)}[n] \right)^{\mathsf{H}}  \mathbf{w}_{i}[n] \right|^{2} \right)  \nonumber \\
&\quad  + \sum\limits_{ i = 1, i \neq m }^{ \mathcal{M} } \left( \frac{ h_0 \| \mathbf{w}_{i}[n] \|^{2} }{ \kappa_{m}^{(l)}[n] + 1 } \right) + \frac{ h_0 }{ \kappa_{m}^{(l)}[n] + 1 } \mathsf{tr}\left(\mathbf{R}_{0}[n] \right) \nonumber \\
&\quad +  \frac{ \kappa_{m}^{(l)}[n] h_0 }{ \kappa_{m}^{(l)}[n] + 1 } \left( \mathbf{\overline{h}}_{m}^{(l)}[n] \right)^{\mathsf{H}} \mathbf{R}_{0}[n]
\mathbf{\overline{h}}_{m}^{(l)}[n] ,
\end{align}
\begin{align}
A^{(l)}[n] & = \frac{1}{2 \sigma_{r}^{2}} \vert \varsigma_{\rm{T}} \vert^2 \tilde{L}  \left(\frac{2 \pi}{\lambda} \cos (\theta_{\rm{T}}^{(l)}[n]) \right)^{2} \left(\mathbf{a}^{(l)}[n]\right)^{\mathsf{H}} \mathbf{R}_{\mathbf{x}}[n] \nonumber \\
& \quad \times \mathbf{a}^{(l)}[n] \mathbf{y}[n]^{\mathsf{T}} \left( \mathbf{I}_{N_{r}} - \frac{1}{N_{r}} \mathbf{1}_{N_{r}} \mathbf{1}_{N_{r}}^{\mathsf{T}} \right) \mathbf{y}[n].
\end{align}

Next, we address the non-concavity of the new objective function $\mathcal{F}_{1}^{(l)}$. For \eqref{p2b}, it can be rewritten as
\begin{align}
\tilde{R}_{m}^{(l)}[n] & = \underbrace{\log_{2} \left( d_{m}^{-2}[n] \left( B_{m}^{(l)}[n] + C_{m}^{(l)}[n] \right) + \sigma^2_{m} \right)}_{R_{m}^{\mathrm{I}}[n]} \nonumber \\
& \quad -  \underbrace{\log_{2} \left(  d_{m}^{-2}[n] C_{m}^{(l)}[n] + \sigma^2_{m} \right)}_{R_{m}^{\mathrm{II}}[n]}.
\end{align}
Clearly, $R_{m}^{\mathrm{I}}[n]$ is a non-convex function with respect to $\mathbf{q}[n]$, but it is convex with respect to $d_{m}^{2}[n]$. Based on the first-order Taylor expansion of a convex function, which serves as a lower bound for the original convex function, we can obtain a lower bound for $R_{m}^{\mathrm{I}}[n]$, given by
\begin{align}
R_{m}^{\mathrm{I}}[n] &\ge 
\left(R_{m}^{\mathrm{I}}[n]\right)^{(l)} 
+ \left(\nabla R_{m}^{\mathrm{I}}[n]\right)^{(l)}
\left( d_{m}^{2}[n] - (d_{m}^{(l)}[n])^{2} \right), \label{p2d}
\end{align}
where 
\begin{align}
&\left(R_{m}^{\mathrm{I}}[n]\right)^{(l)} = \log_{2} \left( \left(d_{m}^{(l)}[n]\right)^{-2} \left( B_{m}^{(l)}[n] + C_{m}^{(l)}[n] \right) + \sigma^2_{m} \right), \\
&\left(\nabla R_{m}^{\mathrm{I}}[n]\right)^{(l)} = - \frac{ \log_{2}({\mathrm{e}}) \left(d_{m}^{(l)}[n] \right)^{-4} \left(B_{m}^{(l)}[n] + C_{m}^{(l)}[n]\right) }{ \left(d_{m}^{(l)}[n] \right)^{-2} \left(B_{m}^{(l)}[n] + C_{m}^{(l)}[n]\right) + \sigma^2_{m} }.
\end{align}

Furthermore, since $R_{m}^{\mathrm{II}}[n]$ is a non-convex function with respect to $\mathbf{q}[n]$, we can introduce an auxiliary variable $\eta_{m}[n]$ to construct a surrogate function that satisfies the upper-bound property of $R_{m}^{\mathrm{II}}[n]$, i.e.,
\begin{equation}
R_{m}^{\mathrm{II}}[n] \le
\log_2 \left(  C_{m}^{(l)}[n] {\mathrm{e}}^{\eta_{m}[n]}   + \sigma^2_{m}  \right), \label{p2e}
\end{equation}
where $\eta_{m}[n]$ satisfies
\begin{equation}
{\mathrm{e}}^{-\eta_{m}[n]} \le \left(d_{m}^{(l)}[n] \right)^{2} + 2\left(\mathbf{q}^{(l)}[n] - \mathbf{q}_{m}\right) \left(\mathbf{q}[n] - \mathbf{q}^{(l)}[n]\right)^{\mathsf{T}}. \label{p2f}
\end{equation}

By substituting \eqref{p2d} and \eqref{p2e} into \eqref{p2b}, we obtain a lower bound for $\tilde{R}_{m}^{(l)}[n]$ as a function of the UAV trajectory $\mathbf{q}[n]$, which is given by
\begin{align}
\tilde{R}_{m}^{(l)}[n] & \geq \left(R_{m}^{\mathrm{I}}[n]\right)^{(l)} 
+ \left(\nabla R_{m}^{\mathrm{I}}[n]\right)^{(l)}
\left( d_{m}^{2}[n] - \left(d_{m}^{(l)}[n]\right)^{2} \right) \nonumber \\
& \quad - \log_2 \left(  C_{m}^{(l)}[n] {\mathrm{e}}^{\eta_{m}[n]}   + \sigma^2_{m}  \right) \nonumber \\
& = \tilde{R}_{m}^{{\rm{lb}},(l)}[n].
\end{align}
For \eqref{p2c}, we can get the lower bound for $\frac{1}{\text{CRB}^{(l)}(\theta_{\rm{T}}[n])}$ as a function of the UAV trajectory $\mathbf{q}[n]$. Specifically, the lower bound is given by
\begin{align}
\frac{1}{\text{CRB}^{(l)}(\theta_{\rm{T}}[n])} 
&\geq A^{(l)}[n] \left( \left( d_{\rm{T}}^{(l)}[n] \right)^{-2} \right.\nonumber  \\
&\quad \left. - \left( d_{\rm{T}}^{(l)}[n] \right)^{-4}\left( d_{\rm{T}}^{2}[n] - \left(d_{\rm{T}}^{(l)}[n]\right)^{2} \right) \right)  \nonumber \\
& = \frac{1}{\text{CRB}^{{\rm{lb}},(l)}(\theta_{\rm{T}}[n])}.
\end{align}

With the above transformations, problem (\textbf{P5}) can be reformulated as the following problem in the $(l+1)$-th iteration of successive convex approximation (SCA),
\begin{subequations} 
\begin{flalign}
 (\textbf{P5-1}.(l+1)):\ & \max_{ \mathbf{q}[n] } \quad \mathcal{F}_{1}^{{\rm{lb}},(l)}\\
 {\rm{s.t.}}  \quad &\eqref{p1b}, \eqref{p1c},\text{ and } \eqref{p2f}, \nonumber
\end{flalign}
\end{subequations}
where 
\begin{equation}
\mathcal{F}_{1}^{{\rm{lb}},(l)} = \xi_{c}  \sum_{n=1}^{N} \sum_{m=1}^{M} \tilde{R}_{m}^{{\rm{lb}},(l)}[n]  + \xi_{s} \sum_{n=1}^{N} \frac{1}{\text{CRB}^{{\rm{lb}},(l)}(\theta_{\rm{T}}[n])}.  
\end{equation}
Obviously, problem $\textbf{P5-1}.(l+1)$ is a standard convex optimization problem, which can be efficiently solved via conventional optimizer, e.g., the CVX toolbox.

\subsection{Overall Algorithm}
The AO-based algorithm for solving problem (\textbf{P1}) is outlined in Algorithm~\ref{algorithm1}, which iteratively alternates among the optimization of the UAV trajectory, the beamforming design, the transmit FA positions, an the receive FA positions until the objective function value converges.

\begin{algorithm}[t] \small
\caption{AO-based Algorithm for Solving Problem (\textbf{P1}).} 
\label{algorithm1} 
\begin{algorithmic}[1]
\STATE
\textbf{Initialize:} $\{\mathbf{q}^{(0)}[n]\}$, $\{ \mathbf{w}_{m}^{(0)}[n], \mathbf{R}_{0}^{(0)}[n] \}$, $ \{\mathbf{x}^{(0)}[n] \}$ and $ \{\mathbf{y}^{(0)}[n] \}$, iteration index $ l = 1 $, maximum iteration number $l_{\rm{max}}$ and accuracy threshold $ \varepsilon > 0 $.
\STATE
\textbf{Repeat:}
\STATE Update auxiliary variables $\omega_{m}[n]$ and $\varpi_{m}[n]$ according to \eqref{auxiliary_variable}. 
\STATE With obtained $\{\mathbf{q}^{(l)}[n]\}$, $ \{\mathbf{x}^{(l)}[n] \}$ and $ \{\mathbf{y}^{(l)}[n] \}$, solve the problem (\textbf{P2-2}) to obtain $\{ \mathbf{w}_{m}^{(l+1)}[n], \mathbf{R}_{0}^{(l+1)}[n] \}$.
\STATE Update auxiliary variables $\omega_{m}[n]$ and $\varpi_{m}[n]$ according to \eqref{auxiliary_variable}. 
\STATE With obtained $\{\mathbf{q}^{(l)}[n]\}$, $\{ \mathbf{w}_{m}^{(l+1)}[n], \mathbf{R}_{0}^{(l+1)}[n] \}$ and $ \{\mathbf{y}^{(l)}[n] \}$, solve the problem (\textbf{P3-2}) to obtain $ \{\mathbf{x}^{(l+1)}[n] \}$.
\STATE With obtained $\{\mathbf{q}^{(l)}[n]\}$, $\{ \mathbf{w}_{m}^{(l+1)}[n], \mathbf{R}_{0}^{(l+1)}[n] \}$ and $ \{\mathbf{x}^{(l+1)}[n] \}$,  update $ \{\mathbf{y}^{(l+1)}[n] \}$ according to \eqref{receive array solution}.
\STATE With obtained $\{ \mathbf{w}_{m}^{(l+1)}[n], \mathbf{R}_{0}^{(l+1)}[n] \}$, $ \{\mathbf{x}^{(l+1)}[n] \}$ and $ \{\mathbf{y}^{(l+1)}[n] \}$, solve the problem (\textbf{P5-1}.($l$)) to obtain $\{\mathbf{q}^{(l+1)}[n]\}$.
\STATE Update the objective function value $\mathcal{F}^{(l)}_{1}$ according to above variables.
\STATE Update $ l = l + 1 $.
\STATE
\textbf{Until:} the increase of the value of the objective function between two
adjacent iterations is smaller than $ \varepsilon $ or $l > l_{\max}$.
\end{algorithmic} 
\end{algorithm}

\subsubsection{Initialization Analysis}
For algorithm initialization, the UAV trajectory is set as a straight-line trajectory between the initial and final locations. The FA elements are uniformly distributed within the feasible region $[0, D_{\mathrm{FA}}]$ with spacing $\frac{D_{\mathrm{FA}}}{K-1}$. The transmit beamforming vectors are initialized using the maximum ratio transmission (MRT) scheme \cite{11082461}.

\subsubsection{Convergence Analysis}
Since the optimal solutions of problems (\textbf{P5-1}.($l$)), (\textbf{P2-2}), and (\textbf{P3-2}) only provide a sub-optimal solution to problem (\textbf{P1}), analyzing the convergence of the AO-based algorithm is essential. The objective function is defined as $\mathcal{F}_{1}\left( \mathbf{q}^{(l)}[n], \mathbf{w}_{m}^{(l)}[n], \mathbf{R}_{0}^{(l)}[n], \mathbf{x}^{(l)}[n], \mathbf{y}^{(l)}[n]  \right)$. In the step 4 of Algorithm \ref{algorithm1}, $\{ \mathbf{w}_{m}^{(l+1)}[n], \mathbf{R}_{0}^{(l+1)}[n] \}$ can be obtained for given $\{\mathbf{q}^{(l)}[n]\}$, $ \{\mathbf{x}^{(l)}[n] \}$ and $ \{\mathbf{y}^{(l)}[n] \}$. Thus, we have
\begin{align}
& \mathcal{F}_{1}\left( \mathbf{q}^{(l)}[n], \mathbf{w}_{m}^{(l)}[n], \mathbf{R}_{0}^{(l)}[n], \mathbf{x}^{(l)}[n], \mathbf{y}^{(l)}[n]  \right) \nonumber \\
& \leq \mathcal{F}_{1}\left( \mathbf{q}^{(l)}[n], \mathbf{w}_{m}^{(l+1)}[n], \mathbf{R}_{0}^{(l+1)}[n], \mathbf{x}^{(l)}[n], \mathbf{y}^{(l)}[n]  \right).
\end{align}
Similarly, from the steps 6-8 of Algorithm \ref{algorithm1}, we can obtain
\begin{align}
& \mathcal{F}_{1}\left( \mathbf{q}^{(l)}[n], \mathbf{w}_{m}^{(l+1)}[n], \mathbf{R}_{0}^{(l+1)}[n], \mathbf{x}^{(l)}[n], \mathbf{y}^{(l)}[n]  \right) \nonumber \\
& \leq \mathcal{F}_{1}\left( \mathbf{q}^{(l)}[n], \mathbf{w}_{m}^{(l+1)}[n], \mathbf{R}_{0}^{(l+1)}[n], \mathbf{x}^{(l+1)}[n], \mathbf{y}^{(l)}[n]  \right) \nonumber \\
& \leq \mathcal{F}_{1}\left( \mathbf{q}^{(l)}[n], \mathbf{w}_{m}^{(l+1)}[n], \mathbf{R}_{0}^{(l+1)}[n], \mathbf{x}^{(l+1)}[n], \mathbf{y}^{(l+1)}[n]  \right) \nonumber \\
& \leq \mathcal{F}_{1}\left( \mathbf{q}^{(l+1)}[n], \mathbf{w}_{m}^{(l+1)}[n], \mathbf{R}_{0}^{(l+1)}[n], \mathbf{x}^{(l+1)}[n], \mathbf{y}^{(l+1)}[n]  \right),
\end{align}
which shows that the objective function value is non-decreasing over iterations. Moreover, the objective function value is upper-bounded due to the limited transmit power. Therefore, Algorithm \ref{algorithm1} is guaranteed to converge.

\subsubsection{Complexity Analysis}
In Algorithm \ref{algorithm1}, subproblems (\textbf{P5-1}.($l$)), (\textbf{P2-2}), and (\textbf{P3-2}) are solved using the interior point method \cite{deng2023beamforming}. The computational complexities of these subproblems are $\mathcal{O}\big((2N)^{3.5} \log(\varepsilon^{-1})\big)$, $\mathcal{O}\big((MNK^{2} + NK^{2})^{3.5} \log(\varepsilon^{-1})\big)$, and $\mathcal{O}\big((K\tilde{I})^{3.5} \log(\varepsilon^{-1})\big)$, respectively. Since the solution to subproblem (\textbf{P4}) can be directly obtained via \eqref{receive array solution}, it does not involve iterative numerical computations, and its computational complexity can therefore be considered negligible. Accordingly, the overall worst-case computational complexity of Algorithm \ref{algorithm1} can be expressed as $\mathcal{O}( l_{\rm{max}}  ( (2N)^{3.5} + (MNK^{2} + NK^{2})^{3.5} + (K\tilde{I})^{3.5}  )\log(1/\epsilon)  )$.

\section{Numerical Results}
This section presents numerical results to demonstrate the efficiency of FAs in the UAV-enabled ISAC system.
As shown in Fig.~\ref{UAV trajectory}, we consider an $800~\rm{m} \times 800~\rm{m}$ square area containing $M = 6$ users and a single-point target, all spatially distributed within the region. Unless otherwise specified, the remaining system parameters are summarized in Table~\ref{tab:parameters}.

\begin{table}[t]
\centering
\caption{Simulation Parameters.}
\label{tab:parameters}
\begin{tabular}{p{6.2cm}c}
\toprule
\centering
Parameters & Value\\
\midrule  
\centering Mission period, $T$ &  45 s\\
\centering Number of time slot, $N$ & 20\\
\centering Number of time interval, $\tilde{I}$ & 5\\
\centering Duration of time slot, $\tau$ & 2.25 s\\
\centering Number of transmit/receive antennas at UAV, $N_{t}/N_{r}$ & 12\\
\centering Channel power gain at reference distance $1 \mathrm{m}$, $h_{0}$& -30 dB\\
\centering Noise power at user receivers, $\sigma_{m}^{2}$& -90 dBm\\
\centering Noise power at UAV receiver, $\sigma_{r}^{2}$& -90 dBm\\
\centering RCS of target \cite{10382465}, $\varsigma_{\rm{T}}$ & $10^{-6}$  \\ 
\centering Length of transmission frame \cite{10382465}, $\tilde{L}$& 256 \\
\centering Antenna mobility region, $[0, D_{\rm{FA}}]$ & $ [0, 20 \lambda ]$ \\
\centering Carrier wavelength \cite{11214460}, $\lambda$ & 0.0107 $\mathrm{m}$\\
\centering Minimum inter-antenna distance \cite{ye2025Fluid}, $d_{\rm{min}}$ & $0.5\lambda$ \\
\centering Maximum UAV speed \cite{deng2023beamforming}, $V_{\rm{max}}$& $20 \ {\mathrm{m/s}}$ \\
\centering UAV flight altitude, $H$ & 100 $\mathrm{m}$ \\
\centering Maximum UAV transmit power, $P_{\rm{max}}$ & 30 dBm \\
\bottomrule
\end{tabular}
\end{table}

\begin{figure}[t]
\centering
\includegraphics[width=0.88\linewidth]{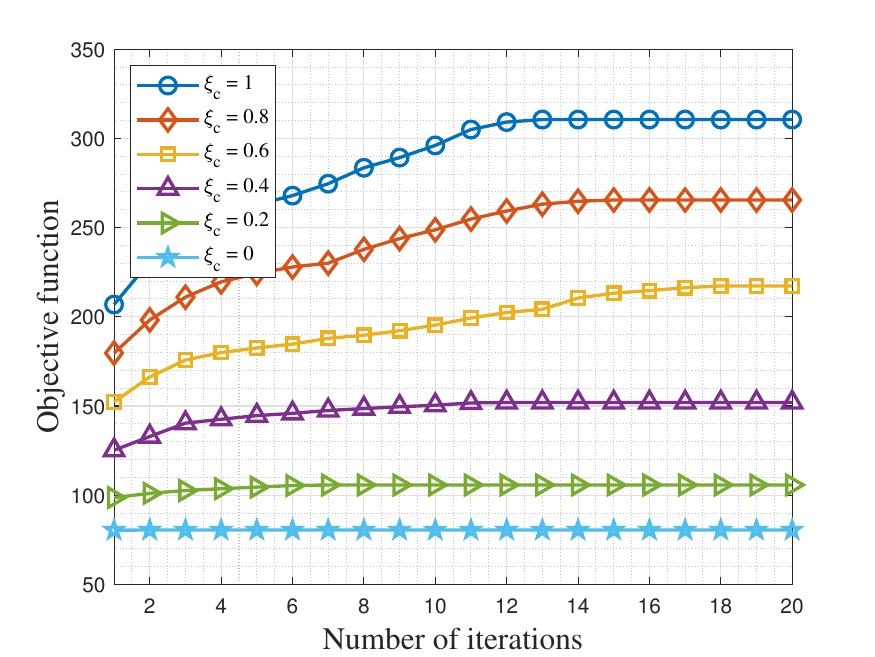}
\caption{The convergence performance of the proposed scheme under different weight factors $\xi_{c}$.}
\label{convergence scheme}
\end{figure}

\begin{figure}[t]
\centering
\includegraphics[width=0.88\linewidth]{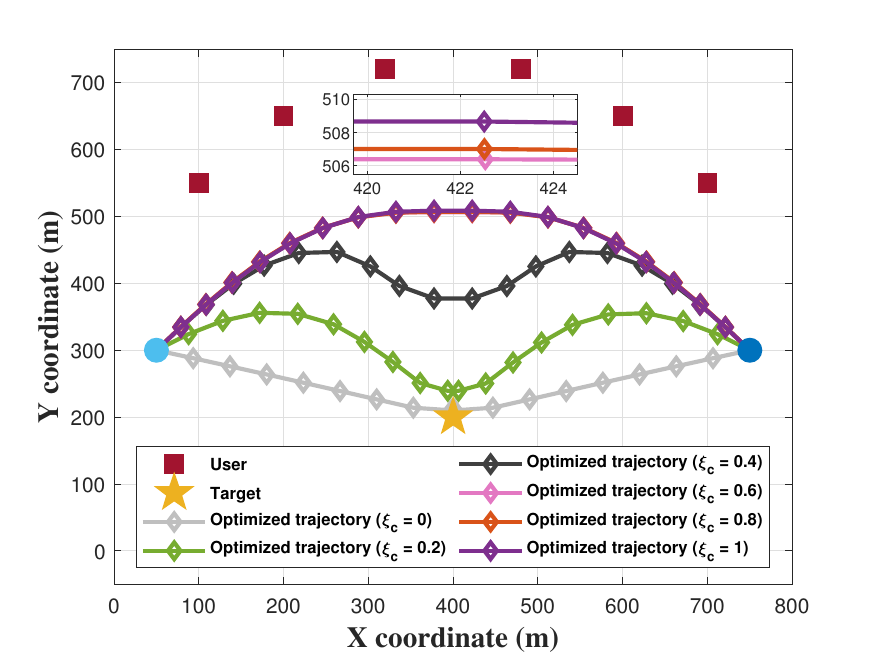}
\caption{The UAV trajectory under different weight factors $\xi_{c}$.}
\label{UAV trajectory}
\end{figure}

\begin{figure}[t]
\centering
\includegraphics[width=0.88\linewidth]{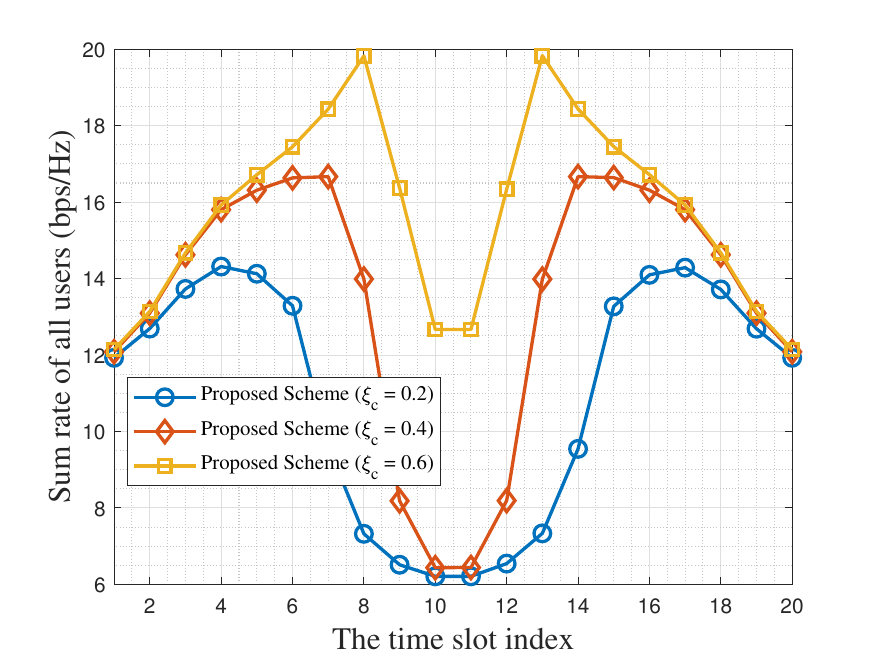}
\caption{The sum rate of all users versus time slot index under different weight factors $\xi_{c}$.}
\label{Rate timeslot}
\end{figure}

\begin{figure}[t]
\centering
\includegraphics[width=0.88\linewidth]{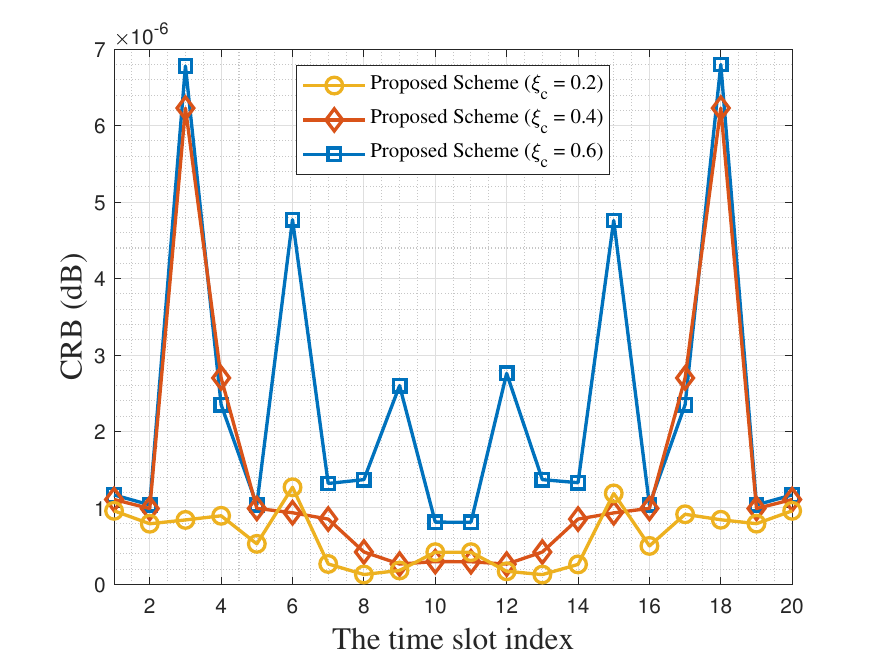}
\caption{The target CRB versus time slot index under different weight factors $\xi_{c}$.}
\label{CRB timeslot}
\end{figure}

\subsection{Performance Analysis of UAV-enabled ISAC System}
First, to comprehensively analyze the trade-off between communication and sensing performance, the system performance is evaluated under different weight factors $\xi_{c}$. Fig. \ref{convergence scheme} presents the convergence behavior of the proposed algorithm under various values of $\xi_{c}$. It is observed that the objective function, defined as the weighted sum of the communication rate and the inverse of the sensing CRB, converges within approximately 15 iterations. When $\xi_{c}=1$, the problem reduces to a communication-only case, achieving the highest objective function value. In contrast, when $\xi_{c}=0$, the optimization becomes a sensing-only case, resulting in the lowest objective function value. For intermediate values of $\xi_{c}$, the proposed algorithm provides a flexible balance between the communication and sensing performance.

As shown in Fig. \ref{UAV trajectory}, the optimized UAV trajectories vary considerably with different weight factors $\xi_{c}$. When $\xi_{c}=0$, the optimization mainly minimizes the sensing CRB, and the UAV tends to approach the target to enhance sensing performance. As $\xi_{c}$ increases, the objective gradually shifts towards improving the communication rate, and the UAV trajectories move closer to the user locations. Finally, when $\xi_{c}=1$, the UAV trajectories are primarily distributed around the user regions to maximize the communication performance.

Figs. \ref{Rate timeslot} and \ref{CRB timeslot} illustrate the variations of the sum rate of all users and the target sensing CRB across the entire flight duration under different weight factors $\xi_{c}$. As shown in Fig. \ref{Rate timeslot}, increasing $\xi_{c}$ notably enhances the communication rate in each time slot, with the highest performance achieved at $\xi_{c}=0.6$. This indicates that the optimization tends to adjust the UAV trajectory toward regions favorable for user communications. However, this improvement comes at the expense of degraded sensing performance, as shown in Fig. \ref{CRB timeslot}, where the CRB values corresponding to $\xi_{c}=0.6$ are generally higher than those under other weight settings. For instance, when $\xi_{c}=0.2$, the system places greater emphasis on sensing accuracy, achieving substantially lower CRB values, but reduced communication rate. A moderate weight factor, such as $\xi_{c}=0.4$, provides a balanced trade-off between the communication and sensing performance. These results demonstrate that adjusting $\xi_{c}$ enables a better trade-off between communication rate and sensing accuracy.

Moreover, when $\xi_{c}=0.2$, the UAV trajectory shifts closer to the target during time slots 7-14 (as shown in Fig. \ref{UAV trajectory}), resulting in the lowest CRB values, which indicates that sensing accuracy is prioritized under this setting. However, during time slots 0-7 and 14-20, the UAV needs to fly toward or away from the target, leading to increased CRB values in these intervals.

\begin{figure*}[t]   
\centering            
\begin{subfigure}{0.32\textwidth}
\includegraphics[width=\linewidth]{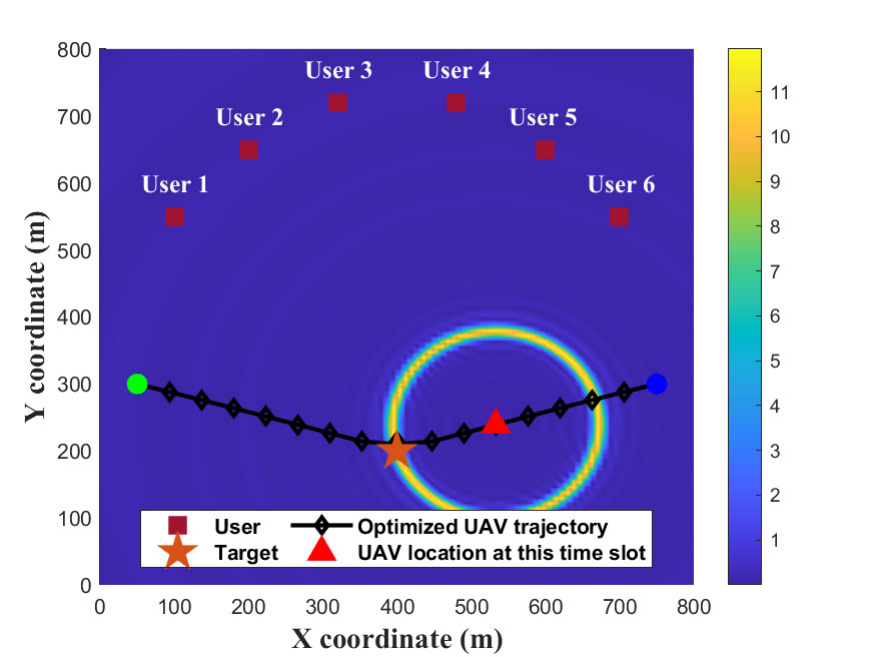}
\caption{Sensing-only scenario.}
\label{fig:beam1}
\end{subfigure}
\hfill
\begin{subfigure}{0.32\textwidth}
\includegraphics[width=\linewidth]{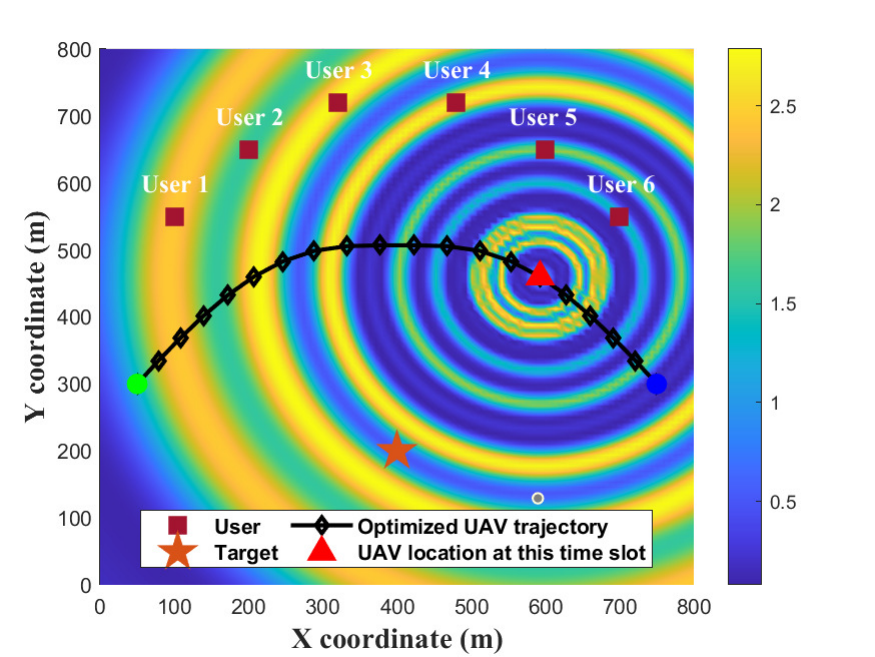}
\caption{Communication-only scenario.}
\label{fig:beam2}
\end{subfigure}
\hfill
\begin{subfigure}{0.32\textwidth}
\includegraphics[width=\linewidth]{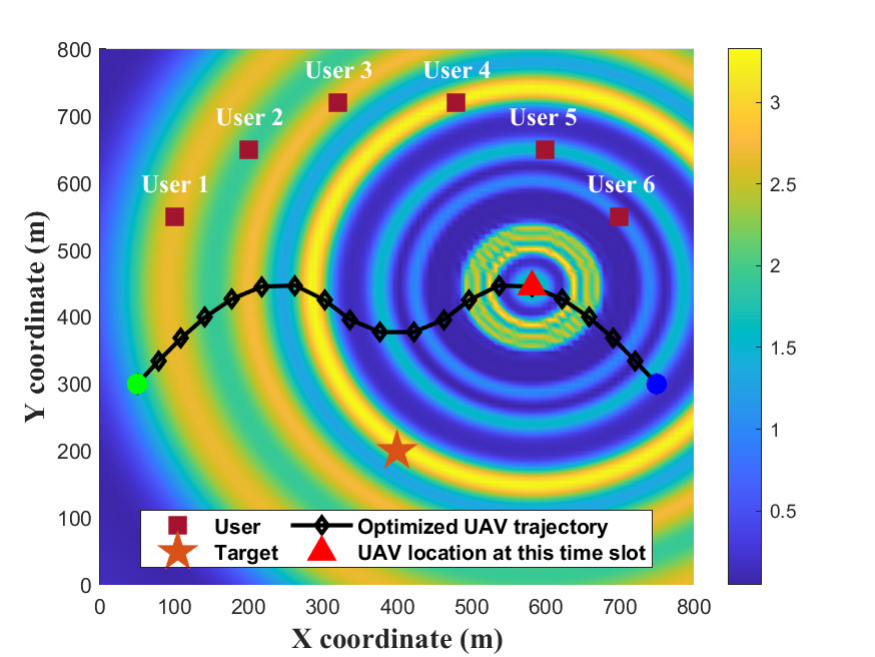}
\caption{Proposed ISAC scenario.}
\label{fig:beam3}
\end{subfigure}
\caption{The achieved beampattern gains in time slot 15 under three different scenarios.}
\label{fig:beamscheme1}            
\end{figure*}

Fig. \ref{fig:beamscheme1} further illustrates the beampattern gain distributions and UAV trajectories under three scenarios: sensing-only, communication-only, and the proposed ISAC. The beampattern gain characterizes the spatial distribution of radiated signal energy and, for any spatial point $(x,y)$, is expressed as
\begin{align}
\mathcal{G}(x,y) = \mathbf{a}^{\mathsf{H}}[n](x,y)\mathbf{R}_{\mathbf{x}}[n] \mathbf{a}[n](x,y),
\end{align}
where $\mathbf{a}^{\mathsf{H}}[n](x,y)$ denotes the steering vector from the UAV to the point $(x,y)$.

As shown in Fig. \ref{fig:beamscheme1}(a), in the sensing-only scenario, the UAV trajectory is biased toward the target location to maximize the reflected signal strength, resulting in a concentrated main lobe around the target and superior sensing gain. However, user coverage is neglected, leading to degraded communication performance when the UAV is far from the users. In contrast, in the communication-only scenario shown in Fig. \ref{fig:beamscheme1}(b), the trajectory is optimized closer to the user locations, resulting in stronger beam energy directed toward the users and improved communication quality, but with significantly reduced sensing gain as the main lobe deviates from the target. Notably, the proposed ISAC scenario in Fig. \ref{fig:beamscheme1}(c) achieves a balanced performance: the UAV trajectory is distributed between the users and the target, while the beampattern provides considerable gain toward both the users and the target. This demonstrates that the proposed scheme effectively coordinates communication and sensing functionalities by optimizing the weighted sum of the communication rate and the sensing CRB.

\begin{figure}[t]
\centering
\includegraphics[width=0.88\linewidth]{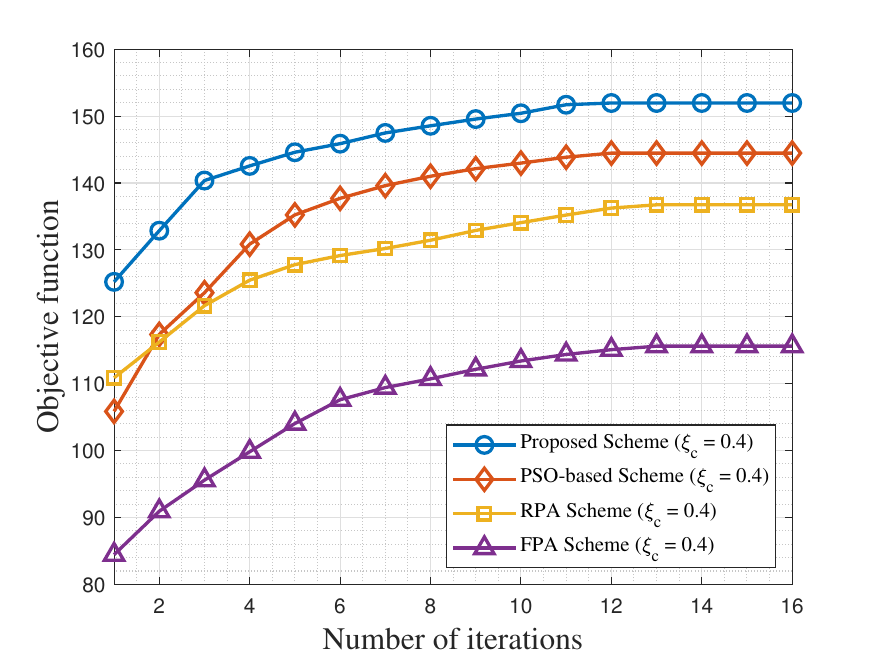}
\caption{The convergence performance under different schemes.}
\label{convergence algorithm}
\end{figure}

\begin{figure}[t]
\centering
\includegraphics[width=0.88\linewidth]{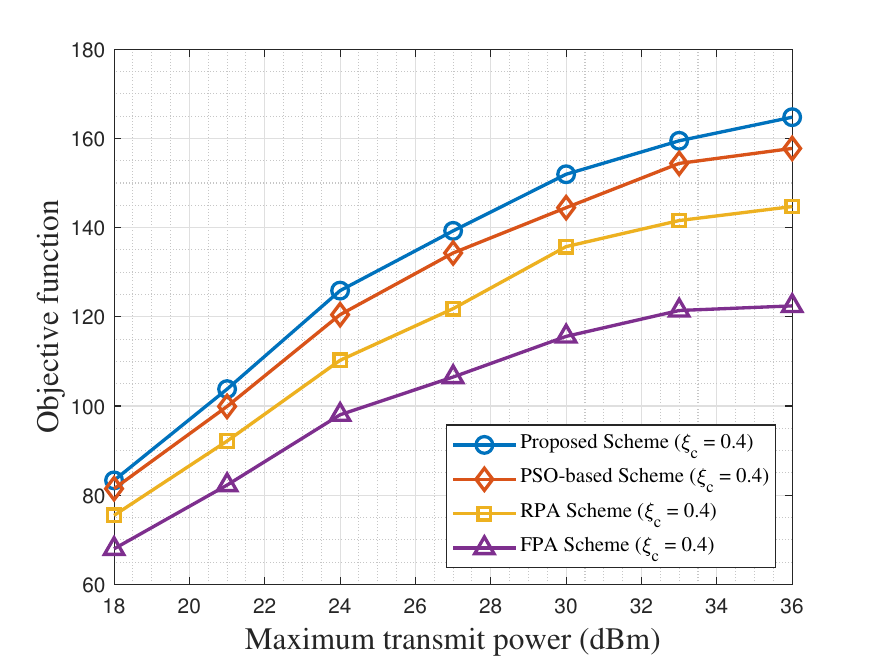}
\caption{The objective function value versus the maximum
 transmit power.}
\label{transmit power}
\end{figure}

\begin{figure}[t]
\centering
\includegraphics[width=0.88\linewidth]{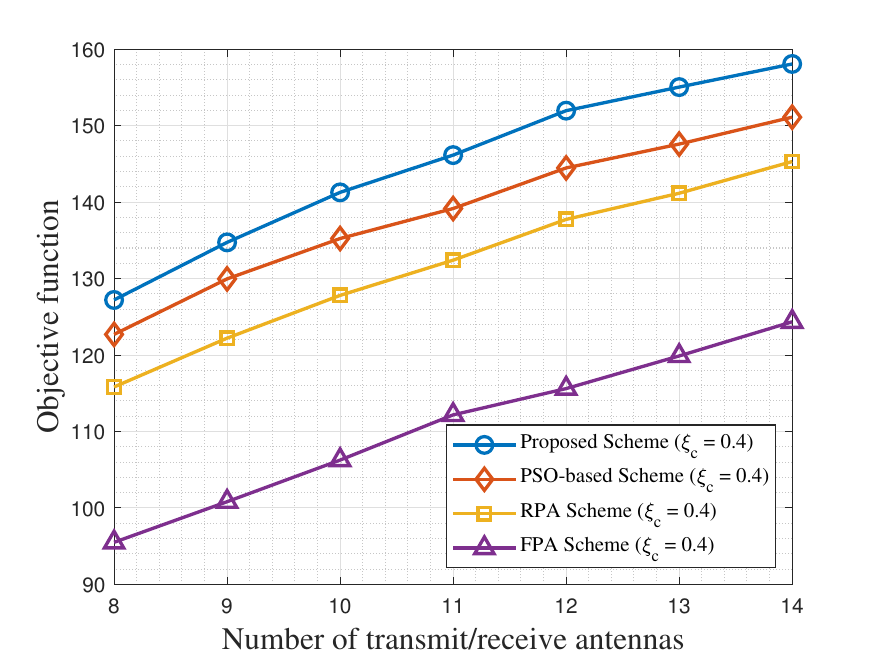}
\caption{The objective function value versus the number of transmit/receive antennas.}
\label{antenna number}
\end{figure}

\begin{figure}[t]
\centering
\includegraphics[width=0.88\linewidth]{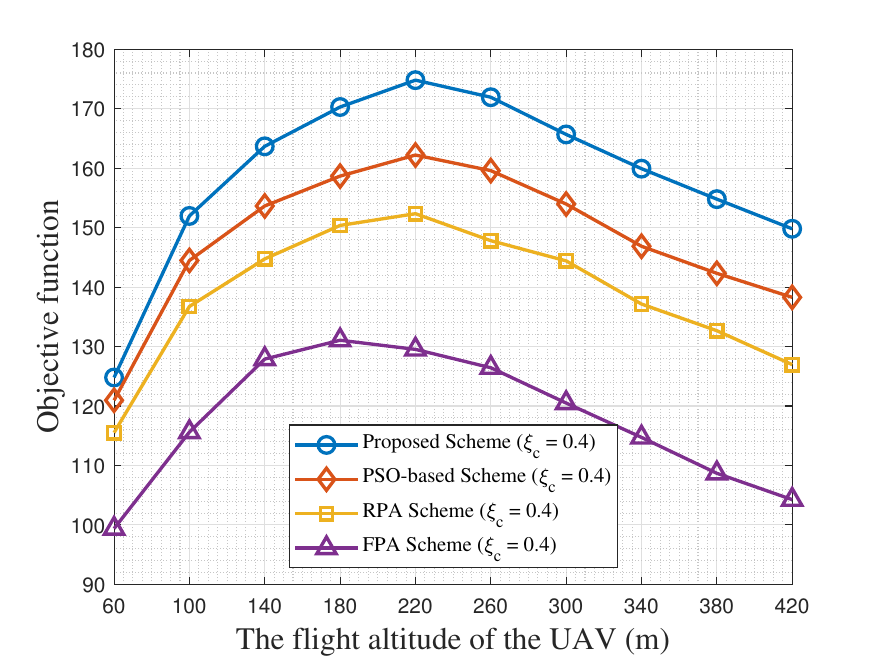}
\caption{The objective function value versus the flight altitude of the UAV.}
\label{UAV flight altitude}
\end{figure}

\subsection{Performance Analysis of FA System}
Next, the performance gains of the FA-based UAV-enabled ISAC system are evaluated by comparing with several benchmark schemes:
\begin{itemize}
\item \textbf{PSO-based Scheme:} The transmit and receive FA positions are optimized using the particle swarm optimization (PSO) algorithm, while all other variables are optimized using the proposed method~\cite{zuo2024Fluid}.
\item \textbf{RPA Scheme:} The transmit and receive FA positions are generated randomly, with the remaining variables optimized using the proposed method.
\item \textbf{FPA Scheme:} The transmit and receive FA positions are configured as ULAs, while all other variables are optimized using the proposed method~\cite{deng2023beamforming}.
\end{itemize}

Fig. \ref{convergence algorithm} compares the convergence behavior of the proposed scheme with the three benchmark algorithms. It is evident that the proposed scheme exhibits a rapid growth of the objective function value in the early iterations and converges to the highest steady-state value after approximately 12 iterations, significantly outperforming all benchmarks. This improvement is attributed to the ability of the proposed method to fully exploit the reconfigurable properties of FA during optimization. The PSO-based scheme, where the transmit and receive FA positions are optimized using PSO, achieves better performance than RPA and FPA owing to its heuristic search capability. However, its final objective function value remains below that of the proposed method.

Fig. \ref{transmit power} illustrates the objective function values achieved by different schemes under varying maximum transmit power $P_{\rm{max}}$. As observed, the objective function values consistently increase with $P_{\rm{max}}$ across all schemes, since higher transmit power enhances the received signal quality and thereby improves both communication and sensing performance. Nevertheless, the performance gaps among the schemes remain evident across the entire power range. The proposed scheme consistently achieves the highest objective function value under all power levels and maintains a significant advantage in the high-power regime.

Fig. \ref{antenna number} presents the objective function values of different schemes versus the number of transmit/receive antennas. It can be observed that the objective function values increase monotonically with the number of antennas for all schemes. This is because a larger array provides higher beamforming gain and greater spatial diversity, which jointly enhance the communication rate and reduce the sensing accuracy. Nevertheless, the performance gap among the schemes remains consistent across the entire antenna range, with the proposed scheme achieving the highest objective function value under all considered antenna configurations.

Fig. \ref{UAV flight altitude} illustrates the impact of UAV flight altitude on the objective function values for different schemes. It is observed that for all schemes, as the altitude increases, the objective function value first rises and then decreases, reaching its peak around 200-250 m. This trend can be explained as follows: a moderate increase in altitude enlarges the elevation angle for both communication and sensing links, thereby improving beam directivity, which benefits both throughput and sensing accuracy. However, when the altitude becomes excessively high, the increased distance leads to higher path loss, which dominates and results in overall performance degradation.

\section{Conclusion}
In this paper, we propose a UAV-enabled ISAC system, in which a single UAV is equipped with transmit and receive FA arrays. By exploiting the mobility of antenna elements, the system provides additional spatial DoFs to enhance both communication and sensing performance. Specifically, a multi-objective optimization problem is formulated to maximize a weighted trade-off between the communication and sensing performance through the joint optimization of the UAV trajectory, beamforming design, and the transmit and receive FA positions. To solve this non-convex problem, an AO-based algorithm is developed to obtain a high-quality solution. Simulation results demonstrate that integrating FA technology into UAV-enabled ISAC systems achieves significantly improved performance compared to conventional FPAs. Future work will focus on extending the framework to multi-UAV cooperative ISAC networks with FAs, incorporating practical FA hardware constraints, and exploring learning-based optimization methods to further enhance real-time adaptability and efficiency.

\begin{appendices}
\section{Detailed Derivation Procedure for \eqref{CRB_2} }\label{proof1}
Based on the definition of the steering vector in \eqref{transmit FA vector} and \eqref{receive FA vector}, we can compute their derivatives with respect to the target vertical angle $\theta_{\rm{T}}[n]$. These derivatives are expressed as
{ \small \begin{align} 
& \dot{\mathbf{a}}[n] \nonumber \\
& = \left[
{\text{j}} \frac{2 \pi}{\lambda} x_{1}[n] \cos (\theta_{\rm{T}}[n]) a_{1}[n],  \ldots, 
{\text{j}} \frac{2 \pi}{\lambda} x_{N_t}[n] \cos (\theta_{\rm{T}}[n]) a_{N_{t}}[n]
\right]^{\mathsf{T}},  \\ 
& \dot{\mathbf{b}}[n] \nonumber \\
& = \left[
{\text{j}} \frac{2 \pi}{\lambda} y_{1}[n] \cos (\theta_{\rm{T}}[n]) b_{1}[n],  \ldots, 
{\text{j}} \frac{2 \pi}{\lambda} y_{N_r}[n] \cos (\theta_{\rm{T}}[n]) b_{N_{r}}[n]
\right]^{\mathsf{T}},
\end{align}}where $ a_{i}[n]$ and $ b_{i}[n]$ denote the $i$-th entries of $\mathbf{a}[n]$ and $\mathbf{b}[n]$, respectively. Using these expressions, it can be easily verified that
{ \begin{align} 
& \dot{\mathbf{a}}^{\mathsf{H}}[n] \mathbf{a}[n] = - {\text{j}} \frac{2 \pi}{\lambda} 
\cos (\theta_{\rm{T}}[n]) \sum_{i = 1}^{N_{t}} x_{i}[n], \label{dot_aa} \\ 
& \dot{\mathbf{b}}^{\mathsf{H}}[n] \mathbf{b}[n] = - {\text{j}} \frac{2 \pi}{\lambda} 
\cos (\theta_{\rm{T}}[n]) \sum_{i = 1}^{N_{r}} y_{i}[n]. \label{dot_bb}
\end{align} }

By utilizing \eqref{covariance_Rx}, \eqref{dot_aa}, and \eqref{dot_bb}, we can derive equations \eqref{CRB_auxiliary1}-\eqref{CRB_auxiliary5}, i.e.,
\begin{align}
 \mathsf{tr}\left(\mathbf{A}^{\mathsf{H}}[n] \mathbf{A}[n] \mathbf{R}_{\mathbf{x}}[n]\right) & =  \mathsf{tr}\left(\mathbf{a}[n] \mathbf{b}^{\mathsf{H}}[n] \mathbf{b}[n] \mathbf{a}^{\mathsf{H}}[n] \mathbf{R}_{\mathbf{x}}[n] \right)  \nonumber \\
& = N_{r} \mathbf{a}^{\mathsf{H}}[n] \mathbf{R}_{\mathbf{x}}[n] \mathbf{a}[n] ,
\label{CRB_auxiliary1}
\end{align}
where $\mathbf{I}_{N_{r}}$ denotes the $N_{r} \times N_{r}$ identity matrix, and $\mathbf{1}_{N_{r}}\mathbf{1}_{N_{r}}^{\mathsf{T}}$ represents the $N_{r} \times N_{r}$ matrix of all ones with $\mathbf{1}_{N_{r}}$ is a column vector of length $N_{r}$ with all entries equal to one.

\begin{figure*}[!t]
\vspace*{-\baselineskip} 
{\small  \begin{align}
\mathsf{tr}\left(\dot{\mathbf{A}}^{\mathsf{H}}[n] \mathbf{A}[n] \mathbf{R}_{\mathbf{x}}[n]\right) &= \mathsf{tr}\left( \left( \mathbf{a}[n] \dot{\mathbf{b}}^{\mathsf{H}}[n] + \dot{\mathbf{a}}[n] \mathbf{b}^{\mathsf{H}}[n] \right) \mathbf{b}[n] \mathbf{a}^{\mathsf{H}}[n] \mathbf{R}_{\mathbf{x}}[n] \right)  \nonumber \\
&= - {\text{j}} \frac{2 \pi}{\lambda} \cos (\theta_{\rm{T}}[n]) \mathbf{a}^{\mathsf{H}}[n] \mathbf{R}_{\mathbf{x}}[n] \mathbf{a}[n] \sum_{i = 1}^{N_{r}} y_{i}[n] + N_{r} \mathbf{a}^{\mathsf{H}}[n] \mathbf{R}_{\mathbf{x}}[n] \dot{\mathbf{a}}[n], \label{CRB_auxiliary2}
\end{align}
\vspace{-\baselineskip}
\begin{align} 
\left| \mathsf{tr}\left( \dot{\mathbf{A}}^{\mathsf{H}}[n] \mathbf{A}[n] \mathbf{R}_{\mathbf{x}}[n] \right) \right|^2 &= \left( - {\text{j}} \frac{2 \pi}{\lambda} \cos (\theta_{\rm{T}}[n]) \mathbf{a}^{\mathsf{H}}[n] \mathbf{R}_{\mathbf{x}}[n] \mathbf{a}[n] \sum_{i = 1}^{N_{r}} y_{i}[n]  +  N_{r} \mathbf{a}^{\mathsf{H}}[n] \mathbf{R}_{\mathbf{x}}[n] \dot{\mathbf{a}}[n] \right)  \nonumber \\
&\quad \times \left( {\text{j}} \frac{2 \pi}{\lambda} \cos (\theta_{\rm{T}}[n]) \mathbf{a}^{\mathsf{H}}[n] \mathbf{R}_{\mathbf{x}}[n] \mathbf{a}[n] \sum_{i = 1}^{N_{r}} y_{i}[n] + N_{r} \dot{\mathbf{a}}^{\mathsf{H}}[n] \mathbf{R}_{\mathbf{x}}[n] \mathbf{a}[n] \right) \nonumber \\
&= \left(\frac{2 \pi}{\lambda} \cos (\theta_{\rm{T}}[n] ) \mathbf{a}^{\mathsf{H}}[n] \mathbf{R}_{\mathbf{x}}[n] \mathbf{a}[n] \right)^{2} \left( \sum_{i = 1}^{N_{r}} y_{i}[n] \right)^{2} + N_{r}^{2} \mathbf{a}^{\mathsf{H}}[n] \mathbf{R}_{\mathbf{x}}[n] \dot{\mathbf{a}}[n] \dot{\mathbf{a}}^{\mathsf{H}}[n] \mathbf{R}_{\mathbf{x}}[n] \mathbf{a}[n]\nonumber \\
&\quad + {\text{j}} \frac{2 \pi}{\lambda} \cos (\theta_{\rm{T}}[n]) N_{r} \mathbf{a}^{\mathsf{H}}[n] \mathbf{R}_{\mathbf{x}}[n] \mathbf{a}[n]
\left( \mathbf{a}^{\mathsf{H}}[n] \mathbf{R}_{\mathbf{x}}[n] \dot{\mathbf{a}}[n] -  \dot{\mathbf{a}}^{\mathsf{H}}[n] \mathbf{R}_{\mathbf{x}}[n] \mathbf{a}[n] \right) \sum_{i = 1}^{N_{r}} y_{i}[n],  \label{CRB_auxiliary3}
\end{align} 
\vspace{-\baselineskip}
\begin{align}
\mathsf{tr}\left(\dot{\mathbf{A}}^{\mathsf{H}}[n] \dot{\mathbf{A}}[n] \mathbf{R}_{\mathbf{x}}[n]\right) & = \mathsf{tr}\left( \left( \mathbf{a}[n] \dot{\mathbf{b}}^{\mathsf{H}}[n] + \dot{\mathbf{a}}[n] \mathbf{b}^{\mathsf{H}}[n] \right) \left( \dot{\mathbf{b}}[n] \mathbf{a}^{\mathsf{H}}[n] + \mathbf{b}[n] \dot{\mathbf{a}}^{\mathsf{H}}[n] \right) \mathbf{R}_{\mathbf{x}}[n]\right) \nonumber \\
& =  \left(\frac{2 \pi}{\lambda} \cos (\theta_{\rm{T}}[n]) \right)^{2} \mathbf{a}^{\mathsf{H}}[n] \mathbf{R}_{\mathbf{x}}[n] \mathbf{a}[n] \sum_{i = 1}^{N_{r}} y_{i}^{2}[n] + N_{r} \dot{\mathbf{a}}^{\mathsf{H}}[n] \mathbf{R}_{\mathbf{x}}[n] \dot{\mathbf{a}}[n]  \nonumber \\
&\quad+ {\text{j}} \frac{2 \pi}{\lambda} 
\cos (\theta_{\rm{T}}[n]) \left( \mathbf{a}^{\mathsf{H}}[n] \mathbf{R}_{\mathbf{x}}[n] \dot{\mathbf{a}}[n] - \dot{\mathbf{a}}^{\mathsf{H}}[n] \mathbf{R}_{\mathbf{x}}[n] \mathbf{a}[n] \right) \sum_{i = 1}^{N_{r}} y_{i}[n].
\label{CRB_auxiliary4}
\end{align}
\vspace{-\baselineskip}
\begin{align}
&\mathsf{tr}\left( \dot{\mathbf{A}}^{\mathsf{H}}[n] \dot{\mathbf{A}}[n] \mathbf{R}_{\mathbf{x}}[n] \right) \mathsf{tr}\left( \mathbf{A}^{\mathsf{H}}[n] \mathbf{A}[n] \mathbf{R}_{\mathbf{x}}[n] \right) - \left\vert \mathsf{tr}\left( \dot{\mathbf{A}}^{\mathsf{H}}[n] \mathbf{A}[n] \mathbf{R}_{\mathbf{x}}[n] \right) \right\vert^2 \nonumber \\
&= N_{r}\left(\frac{2 \pi}{\lambda} \cos (\theta_{\rm{T}}[n] ) \mathbf{a}^{\mathsf{H}}[n] \mathbf{R}_{\mathbf{x}}[n] \mathbf{a}[n] \right)^{2} \left(  \sum_{i = 1}^{N_{r}} y_{i}^{2}[n] - \frac{1}{N_{r}} \left( \sum_{i = 1}^{N_{r}} y_{i}[n] \right)^{2} \right) \nonumber \\
& = N_{r}\left(\frac{2 \pi}{\lambda} \cos (\theta_{\rm{T}}[n] ) \mathbf{a}^{\mathsf{H}}[n] \mathbf{R}_{\mathbf{x}}[n] \mathbf{a}[n] \right)^{2}  \mathbf{y}[n]^{\mathsf{T}} \left( \mathbf{I}_{N_{r}} - \frac{1}{N_{r}} \mathbf{1}_{N_{r}} \mathbf{1}_{N_{r}}^{\mathsf{T}} \right) \mathbf{y}[n] .
\label{CRB_auxiliary5}
\end{align}
} \hrulefill
\end{figure*}

By substituting \eqref{CRB_auxiliary1} and \eqref{CRB_auxiliary5} into \eqref{CRB_1}, we can obtain \eqref{CRB_2}.

\section{Detailed Derivation Procedure for \eqref{ERate_1} }\label{proof2}
According to \cite{gan2021ris}, $\mathbb{E} \big[ R_{m}[n] \big]$ can be approximated as
\begin{equation}
\tilde{R}_{m}[n] = \log_{2} \left( 1 +  \tilde{\gamma}_{m}[n] \right), \label{ERate_2}
\end{equation}
where 
{\small \begin{align} 
&\tilde{\gamma}_{m}[n] \nonumber \\
&= \frac{ \mathbb{E} \{ \vert \mathbf{h}_{m}^{\mathsf{H}}[n] \mathbf{w}_{m}[n]  \vert^{2} \} }
{ \sum\limits_{ i = 1, i \neq m }^{ \mathcal{M} } 
\mathbb{E} \{ \vert  \mathbf{h}_{m}^{\mathsf{H}}[n] \mathbf{w}_{i}[n] \vert^{2} \}
 + \mathbb{E} \{  \mathbf{h}_{m}^{\mathsf{H}}[n] \mathbf{R}_{0}[n] \mathbf{h}_{m}[n] \}   + \sigma^2_{m}}.
\end{align} }

Based on the definition of expectation, we can derive the following expression
\begin{align}
&\mathbb{E} \{ \vert \mathbf{h}_{m}^{\mathsf{H}}[n] \mathbf{w}_{i}[n]  \vert^{2} \} \nonumber \\
& = \mathbb{E} \left\{ \left| \left(\sqrt{ \zeta^{\rm{los}}_{m}[n] } \mathbf{\overline{h}}_{m}[n] + \sqrt{ \zeta^{\rm{nlos}}_{m}[n] } \mathbf{\tilde{h}}_{m}[n] \right)^{\mathsf{H}} \mathbf{w}_{i}[n]  \right|^{2} \right\} \nonumber \\
& = \zeta^{\rm{los}}_{m}[n] \left\vert \mathbf{\overline{h}}_{m}^{\mathsf{H}}[n]\mathbf{w}_{i}[n] \right\vert^{2} + \zeta^{\rm{nlos}}_{m}[n] \| \mathbf{w}_{i}[n] \|^{2}, \label{ERate_3}
\end{align}
where $\zeta^{\rm{los}}_{m}[n] = \frac{ \kappa_{m}[n] \beta_{m}[n] }{ \kappa_{m}[n] + 1 }$ and $\zeta^{\rm{nlos}}_{m}[n] = \frac{ \beta_{m}[n] }{ \kappa_{m}[n] + 1 }$.
Similarly, we can obtain that
\begin{align}
& \mathbb{E} \{  \mathbf{h}_{m}^{\mathsf{H}}[n] \mathbf{R}_{0}[n] \mathbf{h}_{m}[n] \} \nonumber \\
&= \zeta^{\rm{los}}_{m}[n] \mathbf{\overline{h}}_{m}^{\mathsf{H}}[n] \mathbf{R}_{0}[n] \mathbf{\overline{h}}_{m}[n] + \zeta^{\rm{nlos}}_{m}[n]\mathsf{tr}\left(\mathbf{R}_{0}[n] \right). \label{ERate_4}
\end{align}

By substituting \eqref{ERate_3} and \eqref{ERate_4} into \eqref{ERate_2}, we can obtain \eqref{ERate_1}.

\section{Proof of the Rank-One Property of $\{ \mathbf{W}_{m}^{\rm{opt}}[n] \}$ } \label{proof3}
It is evident that problem (\textbf{P2-2}) constitutes a convex optimization problem satisfying the Slater's condition. Consequently, a zero duality gap exists between the primal and dual problems.
Retaining only relevant terms, the Lagrangian of (\textbf{P2-2}) with respect to transmit beamforming matrices $\{\mathbf{W}_{m}[n]\}$ is
\begin{align}
\mathbf{\mathcal{L}} & = \xi_{\text{c}} \sum_{n=1}^{N} \sum_{m=1}^{M} \tilde{R}^{\rm{III}}_{m}[n] +\xi_{\text{s}} \sum_{n=1}^{N} \frac{1}{\text{CRB}(\theta_{\rm{T}}[n])} \nonumber \\
& \quad + \sum_{n=1}^{N} \lambda_{n} \left( P_{\max} - \sum_{m= 1}^{M} \mathsf{tr}\left( \mathbf{W}_{m}[n] \right) - \mathsf{tr}(\mathbf{R}_{0}[n]) \right) \nonumber \\
& \quad + \sum_{n=1}^{N} \sum_{m=1}^{M} \mathsf{tr}(\mathbf{\Omega}_{m,n}\mathbf{W}_{m}[n]) ,
\end{align}
where $\{ \mathbf{\Omega}_{m,n} \succeq \mathbf{0},\forall m \in \mathcal{M}, \forall n \in \mathcal{N} \}$ denotes the dual matrix for the positive semi-definite constraint \eqref{p3d}, and $\{ \lambda_{n} \geq 0,\forall n \in \mathcal{N} \}$ represents the dual variable for the power constraint \eqref{p3e}. The KKT conditions are then given by
\begin{align}
\dfrac{\partial \mathbf{\mathcal{L}}}{\partial \mathbf{W}_{m}[n]} &= \xi_{\text{c}} \varpi_{m}[n] \sqrt{1 + \omega_{m}[n]}  E_{m}^{-\frac{1}{2}}[n] \mathbf{B}_{m}[n] \nonumber \\
&\quad - \xi_{\text{c}}\sum_{i=1}^{M} \varpi_{i}^{2}[n] \mathbf{B}_{i}[n] + \mathbf{\Omega}_{m,n} - \lambda_{n} \mathbf{I}_{N_{t}} \nonumber \\
& \quad + \xi_{\text{s}} D_{m}[n] \mathbf{a}[n]\mathbf{a}^{\mathsf{H}}[n] , \label{proof3-1}
\end{align}
\begin{equation}
\mathsf{tr}(\mathbf{\Omega}_{m,n}\mathbf{W}_{m}[n]) = 0, \label{proof3-2}
\end{equation}
where $\mathbf{B}_{i}[n] = \zeta^{\mathrm{los}}_{i}[n] \overline{\mathbf{h}}_{i}[n] \overline{\mathbf{h}}_{i}^{\mathrm{H}}[n] + \zeta^{\mathrm{nlos}}_{i}[n] \mathbf{I}_{N_{t}}, \forall i \in \mathcal{M}$.

From \eqref{proof3-1}, we can derive the following expression for the matrix $\mathbf{\Omega}_{m,n}$, and is given by \eqref{proof3-3}.
It is observed that $ \mathbf{\Upsilon} $ is a linear combination of rank-one matrices, and hence satisfies $ \mathsf{rank}(\mathbf{\Upsilon}) \geq 1 $. 
Given that $\mathbf{\Omega}_{m,n} \succeq \mathbf{0}$, it follows that $\varrho \geq 0$ and $\varrho + \tilde{\lambda}_{\mathbf{\Upsilon}} \geq 0$, where $\tilde{\lambda}_{\mathbf{\Upsilon}}$ denotes the eigenvalue of $\mathbf{\Upsilon}$.
Consequently, the matrix $ \mathbf{\Omega}_{m,n} $ can be decomposed as the sum of a scaled identity matrix of full rank and a rank-one matrix. This decomposition implies that $ \mathsf{rank}(\mathbf{\Omega}_{m,n}) \geq N_t - 1 $.

From \eqref{proof3-2}, we have $\mathbf{\Omega}_{m,n} \mathbf{W}_{m}[n]= \mathbf{0}$, which implies that the columns of $ \mathbf{W}_{m}[n] $ lie in the null space of $ \mathbf{\Omega}_{m,n} $. Since $ \mathsf{rank}(\mathbf{\Omega}_{m,n}) \geq N_t - 1 $, the dimension of its null space is at most one. Consequently, the rank of $ \mathbf{W}_{m}[n] $ satisfies $\mathsf{rank}(\mathbf{W}_{m}[n]) \leq N_t - \mathsf{rank}(\mathbf{\Omega}_{m,n}) \leq 1.$ Furthermore, since $ \mathbf{W}_{m}[n] \succeq \mathbf{0} $ and the optimal solution must satisfy $ \mathbf{W}_{m}[n] \neq \mathbf{0} $, it follows that $ \mathsf{rank}(\mathbf{W}_{m}[n]) = 1 $. This completes the proof.

\begin{figure*}[!t]
\vspace*{-\baselineskip} 
{\small \begin{align}
\mathbf{\Omega}_{m,n} & = \lambda_{n} \mathbf{I}_{N_{t}} + \xi_{c}\sum_{i=1}^{M} \varpi_{i}^{2}[n] \mathbf{B}_{i}[n] - \xi_{c} \varpi_{m}[n] \sqrt{1 + \omega_{m}[n]}  E_{m}^{-\frac{1}{2}}[n] \mathbf{B}_{m}[n] \nonumber \\
&=\left( \underbrace{ \lambda_{n} + \xi_{c}\sum_{i=1}^{M} \varpi_{i}^{2}[n] \zeta^{\mathrm{nlos}}_{i}[n] - \xi_{c} \varpi_{m}[n] \sqrt{1 + \omega_{m}[n]}  E_{m}^{-\frac{1}{2}}[n] \zeta^{\mathrm{nlos}}_{m}[n] }_{ \varrho }  \right) \mathbf{I}_{N_{t}} \nonumber \\
&\quad + \underbrace{ \xi_{c}\sum_{i=1}^{M} \varpi_{i}^{2}[n]\zeta^{\mathrm{los}}_{i}[n] \overline{\mathbf{h}}_{i}[n] \overline{\mathbf{h}}_{i}^{\mathrm{H}}[n] - \xi_{c} \varpi_{m}[n] \sqrt{1 + \omega_{m}[n]}  E_{m}^{-\frac{1}{2}}[n]\overline{\mathbf{h}}_{m}[n] \overline{\mathbf{h}}_{m}^{\mathrm{H}}[n] - \xi_{s} D_{m}[n] \mathbf{a}[n]\mathbf{a}^{\mathsf{H}}[n] }_{ \mathbf{\Upsilon} }.
\label{proof3-3}
\end{align}
} \hrulefill
\end{figure*}

\section{Proof of Negative Semi-Definiteness of $\mathbf{S}^{\rm{I}}_{m}[n]$ } \label{proof4}
We first define $\mathbf{1}_{N_{t}} = [1, 1, \dots, 1]^{\mathsf{T}} \in \mathbb{R}^{N_{t} \times 1}$, and introduce the diagonal matrix
\begin{equation}\small
\varkappa_{m}[n] = \mathsf{diag}\left(\sqrt{\vert [ \mathbf{w}_{m}[n] ]_{1} \vert}, \dots, \sqrt{\vert [ \mathbf{w}_{m}[n] ]_{N_{t}} \vert}^{\mathsf{T}}\right) \in \mathbb{R}^{N_{t} \times N_{t}}.
\end{equation}
We also define $\varsigma_{m}[n] \triangleq \sum_{q=1}^{N_{t}} \vert [ \mathbf{w}_{m}[n] ]_{q} \vert = \| \varkappa_{m}[n] \mathbf{1}_{N_{t}} \|_2^2 $.

Using these definitions, the diagonal matrix $\text{diag} \left( \mathbf{V}_{m}^{\rm{I}}[n] \right)$ and the matrix $\mathbf{V}_{m}^{\rm{II}}[n]$ can be rewritten as $\varsigma_{m}[n]\varkappa_{m}[n]\varkappa_{m}[n]$ and $\varkappa_{m}[n]\varkappa_{m}[n]\mathbf{1}_{N_{t}}(\varkappa_{m}[n]\varkappa_{m}[n]\mathbf{1}_{N_{t}})^{\mathsf{T}}$.
Therefore, the quadratic form $\mathbf{x}[n]^{\mathsf{T}} \mathbf{S}^{\rm{I}}_{m}[n] \mathbf{x}[n]$ can be expressed as
\begin{align}
&\mathbf{x}[n]^{\mathsf{T}} \mathbf{S}^{\rm{I}}_{m}[n] \mathbf{x}[n] \nonumber \\
&= -2 \vartheta_{m}^{2}[n]  \mathbf{x}[n]^{\mathsf{T}}\left( \text{diag} \left( \mathbf{V}_{m}^{\rm{I}}[n] \right)-\mathbf{V}_{m}^{\rm{II}}[n]\right)\mathbf{x}[n], \nonumber \\
& = -2 \vartheta_{m}^{2}[n] \left( \varsigma_{m}[n] \| \varkappa_{m}[n]  \mathbf{x}[n] \|^{2} -  \vert \mathbf{1}^{\mathsf{T}}_{N_{t}}  \varkappa_{m}[n]\varkappa_{m}[n] \mathbf{x}[n] \vert^{2}  \right) \nonumber \\
&= -2 \vartheta_{m}^{2}[n] \left( \| \varkappa_{m}[n] \mathbf{1}_{N_{t}} \|_2^2 \| \varkappa_{m}[n]  \mathbf{x}[n] \|^{2} \right. \nonumber \\
& \quad \left. -  \vert \mathbf{1}^{\mathsf{T}}_{N_{t}}  \varkappa_{m}[n]\varkappa_{m}[n] \mathbf{x}[n] \vert^{2}  \right).
\end{align}
By the Cauchy-Schwarz inequality, which states that $\| \mathbf{u} \|_2^2 \| \mathbf{v} \|_2^2 \geq |\mathbf{u}^{\mathsf{T}} \mathbf{v}|^2$ for any two vectors $\mathbf{u}$ and $\mathbf{v}$ of the same dimension, we conclude that $\mathbf{x}[n]^{\mathsf{T}} \mathbf{S}^{\rm{I}}_{m}[n] \mathbf{x}[n] \leq 0$. Hence,  the matrix $\mathbf{S}^{\rm{I}}_{m}[n]$ is negative semi-definite. This completes the proof.

\section{Proof of the Optimal solution for receiving FA position } \label{proof5}
To maximize the TSS of the sequence $\mathbf{y}[n]$, the optimal solution must satisfy the following
\begin{equation}
    y_{1}[n] = 0,\ y_{N_{r}}[n] = D_{\rm{FA}},\ \sum_{i=1}^{N_{r}-1} \delta_{i}[n] = D_{\rm{FA}},
\end{equation}
where $\delta_{i}[n] = y_{i+1}[n] - y_{i}[n] \geq d_{\rm{min}}$ are auxiliary variables.

Therefore, problem (\textbf{P4}) can be reformulated as
\begin{subequations} 
\begin{flalign}
 (\textbf{P4-1}):\ & \max_{ \boldsymbol{\delta}_{i}[n] } \quad \sum_{n=1}^{N} \varphi[n] \left(  \sum_{i = 1}^{N_{r}} y_{i}^{2}[n] - \frac{1}{N_{r}} \left( \sum_{i = 1}^{N_{r}} y_{i}[n] \right)^{2} \right)  \label{p5b} \\
 {\rm{s.t.}}  \quad & \delta_{j}[n] \geq d_{\rm{min}}, j = 1,2,\ldots, N_{r}-1, \forall n \in \mathcal{N}, \\
 & \sum_{j=1}^{N_{r}-1} \delta_{j}[n] = D_{\rm{FA}}, \forall n \in \mathcal{N} , \\
 & \eqref{p1f},\nonumber
\end{flalign}
\end{subequations}
where $y_{i}[n] = \sum_{i=1}^{i-1} \delta_{i}[n], i = 2,3,\ldots, N_{r}$.
It is evident that problem (\textbf{P4-1}) is a convex optimization problem with respect to $\boldsymbol{\delta}_{i}[n]$, and the optimal solution is achieved at the boundary of the feasible region. The boundary satisfies the following condition: $N_{r} - 2$ of the $\delta_i$'s are equal to the minimum distance $d_{\rm{min}}$, while the remaining one, denoted $\delta_j$, is given by $D_{\rm{FA}} - (N_r - 2)d_{\rm{min}}$. 
The resulting sequence is given by
\begin{equation} 
y_{p}[n] = \begin{cases} 
(p-1)d_{\rm{min}}, & p = 1, \dots, j, \\
D_{\rm{FA}} - (N_{r} - p)d_{\rm{min}}, & p = j+1, \dots, N_{r}.
\end{cases}
\end{equation}

Next, we determine the optimal index $j$, which specifies the position in the sequence where the placement shifts from the lower-bound cluster to the upper-bound cluster. If $j$ is too small, the transition occurs prematurely, reducing dispersion in the upper-bound region. Conversely, if $j$ is too large, the transition is delayed, limiting dispersion in the lower-bound region.
Accordingly, the optimal index can be expressed as
\begin{equation} 
j^{\rm{opt}} = \begin{cases} 
\frac{N_{r}}{2}, & \text{if $N_{r}$ is even}, \\
\frac{N_{r}-1}{2}\ \text{or} \ \frac{N_{r}+1}{2}, & \text{if $N_{r}$ is odd}.
\end{cases}
\end{equation}

\end{appendices}

\bibliographystyle{IEEEtran}
\bibliography{myref}
\end{document}